\pdfoutput=1
\documentclass[11pt,a4paper]{article}
\usepackage{jheppub}
\usepackage{microtype}
\usepackage{dcolumn}
\usepackage{booktabs}

\usepackage{lineno}  

\usepackage{xspace}

\def\bfseries{\fontseries \bfdefault \selectfont \boldmath}

\newcommand{\Ba}[1]{\ensuremath{^{#1}\mathrm{Ba}}\xspace}

\newcommand{\Co}[1]{\ensuremath{^{#1}\mathrm{Co}}\xspace}

\newcommand{\K}[1]{\ensuremath{^{#1}\mathrm{K}}\xspace}
\newcommand{\Kr}[1]{\ensuremath{^{#1}\mathrm{Kr}}\xspace}

\newcommand{\Th}[1]{\ensuremath{^{#1}\mathrm{Th}}\xspace}

\newcommand{\U}[1]{\ensuremath{^{#1}\mathrm{U}}\xspace}
\newcommand{\Xe}[1]{\ensuremath{^{#1}\mathrm{Xe}}\xspace}
\newcommand{\Te}[1]{\ensuremath{^{#1}\mathrm{Te}}\xspace}

\newcommand{\bbnonu}{\ensuremath{0\nu\beta\beta}\xspace}
\newcommand{\bbtwonu}{\ensuremath{2\nu\beta\beta}\xspace}
\newcommand{\bb}{\ensuremath{\beta\beta}\xspace}

\newcommand{\halflife}{\ensuremath{T_{1/2}}\xspace}  
\newcommand{\ececnonu}{\ensuremath{0\nu ECEC}\xspace}
\newcommand{\ecectwonu}{\ensuremath{2\nu ECEC}\xspace}
\newcommand{\ecec}{\ensuremath{ECEC}\xspace}

\newcommand{\Sone}{\ensuremath{S_1}\xspace}
\newcommand{\Stwo}{\ensuremath{S_2}\xspace}

\newcommand{\new}{NEXT-White\xspace}
\newcommand{\nexthundred}{NEXT-100\xspace}

\begin{document}

\title{Sensitivity of the NEXT experiment to Xe-124 double electron capture}

\collaboration{The NEXT Collaboration}
\author[19,21,a]{G.~Mart\'inez-Lema,\note[a]{Corresponding author. Now at Weizmann Institute of Science, Israel.}}
\author[19,16]{M.~Mart\'inez-Vara,}
\author[19]{M.~Sorel,}
\author[2]{C.~Adams,}
\author[22]{V.~\'Alvarez,}
\author[6]{L.~Arazi,}
\author[20]{I.J.~Arnquist,}
\author[4]{C.D.R~Azevedo,}
\author[2]{K.~Bailey,}
\author[22]{F.~Ballester,}
\author[16,19]{J.M.~Benlloch-Rodr\'{i}guez,}
\author[14]{F.I.G.M.~Borges,}
\author[3]{N.~Byrnes,}
\author[19]{S.~C\'arcel,}
\author[19]{J.V.~Carri\'on,}
\author[23]{S.~Cebri\'an,}
\author[20]{E.~Church,}
\author[14]{C.A.N.~Conde,}
\author[11]{T.~Contreras,}
\author[21]{G.~D\'iaz,}
\author[19]{J.~D\'iaz,}
\author[5]{M.~Diesburg,}
\author[14]{J.~Escada,}
\author[22]{R.~Esteve,}
\author[6,7,19]{R.~Felkai,}
\author[13]{A.F.M.~Fernandes,}
\author[13]{L.M.P.~Fernandes,}
\author[16,9]{P.~Ferrario,}
\author[4]{A.L.~Ferreira,}
\author[13]{E.D.C.~Freitas,}
\author[16]{J.~Generowicz,}
\author[11]{S.~Ghosh,}
\author[8]{A.~Goldschmidt,}
\author[16,9,b]{J.J.~G\'omez-Cadenas,\note[b]{NEXT Co-spokesperson.}}
\author[21]{D.~Gonz\'alez-D\'iaz,}
\author[11]{R.~Guenette,}
\author[10]{R.M.~Guti\'errez,}
\author[11]{J.~Haefner,}
\author[2]{K.~Hafidi,}
\author[1]{J.~Hauptman,}
\author[13]{C.A.O.~Henriques,}
\author[21]{J.A.~Hernando~Morata,}
\author[16]{P.~Herrero,}
\author[22]{V.~Herrero,}
\author[6,7]{Y.~Ifergan,}
\author[2]{S.~Johnston,}
\author[3]{B.J.P.~Jones,}
\author[21,19]{M.~Kekic,}
\author[18]{L.~Labarga,}
\author[3]{A.~Laing,}
\author[5]{P.~Lebrun,}
\author[22,19]{N.~L\'opez-March,}
\author[10]{M.~Losada,}
\author[13]{R.D.P.~Mano,}
\author[19]{J.~Mart\'in-Albo,}
\author[19,16]{A.~Mart\'inez,}
\author[3]{A.D.~McDonald,}
\author[16,9]{F.~Monrabal,}
\author[13]{C.M.B.~Monteiro,}
\author[22]{F.J.~Mora,}
\author[19,16]{J.~Mu\~noz Vidal,}
\author[19]{P.~Novella,}
\author[3,c]{D.R.~Nygren,\note[c]{NEXT Co-spokesperson.}}
\author[21,19]{B.~Palmeiro,}
\author[5]{A.~Para,}
\author[12]{J.~P\'erez,}
\author[19]{M.~Querol,}
\author[6]{A.B.~Redwine,}
\author[21,19]{J.~Renner,}
\author[2]{J.~Repond,}
\author[2]{S.~Riordan,}
\author[17]{L.~Ripoll,}
\author[10]{Y.~Rodr\'iguez Garc\'ia,}
\author[22]{J.~Rodr\'iguez,}
\author[3]{L.~Rogers,}
\author[16,12]{B.~Romeo,}
\author[19]{C.~Romo-Luque,}
\author[14]{F.P.~Santos,}
\author[13]{J.M.F. dos~Santos,}
\author[6]{A.~Sim\'on,}
\author[15,d]{C.~Sofka,\note[d]{Now at University of Texas at Austin, USA.}}
\author[15]{T.~Stiegler,}
\author[22]{J.F.~Toledo,}
\author[16]{J.~Torrent,}
\author[19]{A.~Us\'on,}
\author[4]{J.F.C.A.~Veloso,}
\author[15]{R.~Webb,}
\author[6,e]{R.~Weiss-Babai,\note[e]{On leave from Soreq Nuclear Research Center, Yavneh, Israel.}}
\author[15,f]{J.T.~White,\note[f]{Deceased.}}
\author[3]{K.~Woodruff,}
\author[19]{N.~Yahlali}
\emailAdd{gonzalo.martinez.lema@weizmann.ac.il}
\affiliation[1]{
Department of Physics and Astronomy, Iowa State University, 12 Physics Hall, Ames, IA 50011-3160, USA}
\affiliation[2]{
Argonne National Laboratory, Argonne, IL 60439, USA}
\affiliation[3]{
Department of Physics, University of Texas at Arlington, Arlington, TX 76019, USA}
\affiliation[4]{
Institute of Nanostructures, Nanomodelling and Nanofabrication (i3N), Universidade de Aveiro, Campus de Santiago, Aveiro, 3810-193, Portugal}
\affiliation[5]{
Fermi National Accelerator Laboratory, Batavia, IL 60510, USA}
\affiliation[6]{
Nuclear Engineering Unit, Faculty of Engineering Sciences, Ben-Gurion University of the Negev, P.O.B. 653, Beer-Sheva, 8410501, Israel}
\affiliation[7]{
Nuclear Research Center Negev, Beer-Sheva, 84190, Israel}
\affiliation[8]{
Lawrence Berkeley National Laboratory (LBNL), 1 Cyclotron Road, Berkeley, CA 94720, USA}
\affiliation[9]{
Ikerbasque, Basque Foundation for Science, Bilbao, E-48013, Spain}
\affiliation[10]{
Centro de Investigaci\'on en Ciencias B\'asicas y Aplicadas, Universidad Antonio Nari\~no, Sede Circunvalar, Carretera 3 Este No.\ 47 A-15, Bogot\'a, Colombia}
\affiliation[11]{
Department of Physics, Harvard University, Cambridge, MA 02138, USA}
\affiliation[12]{
Laboratorio Subterr\'aneo de Canfranc, Paseo de los Ayerbe s/n, Canfranc Estaci\'on, E-22880, Spain}
\affiliation[13]{
LIBPhys, Physics Department, University of Coimbra, Rua Larga, Coimbra, 3004-516, Portugal}
\affiliation[14]{
LIP, Department of Physics, University of Coimbra, Coimbra, 3004-516, Portugal}
\affiliation[15]{
Department of Physics and Astronomy, Texas A\&M University, College Station, TX 77843-4242, USA}
\affiliation[16]{
Donostia International Physics Center (DIPC), Paseo Manuel Lardizabal, 4, Donostia-San Sebastian, E-20018, Spain}
\affiliation[17]{
Escola Polit\`ecnica Superior, Universitat de Girona, Av.~Montilivi, s/n, Girona, E-17071, Spain}
\affiliation[18]{
Departamento de F\'isica Te\'orica, Universidad Aut\'onoma de Madrid, Campus de Cantoblanco, Madrid, E-28049, Spain}
\affiliation[19]{
Instituto de F\'isica Corpuscular (IFIC), CSIC \& Universitat de Val\`encia, Calle Catedr\'atico Jos\'e Beltr\'an, 2, Paterna, E-46980, Spain}
\affiliation[20]{
Pacific Northwest National Laboratory (PNNL), Richland, WA 99352, USA}
\affiliation[21]{
Instituto Gallego de F\'isica de Altas Energ\'ias, Univ.\ de Santiago de Compostela, Campus sur, R\'ua Xos\'e Mar\'ia Su\'arez N\'u\~nez, s/n, Santiago de Compostela, E-15782, Spain}
\affiliation[22]{
Instituto de Instrumentaci\'on para Imagen Molecular (I3M), Centro Mixto CSIC - Universitat Polit\`ecnica de Val\`encia, Camino de Vera s/n, Valencia, E-46022, Spain}
\affiliation[23]{
Centro de Astropart\'iculas y F\'isica de Altas Energ\'ias (CAPA), Universidad de Zaragoza, Calle Pedro Cerbuna, 12, Zaragoza, E-50009, Spain}

\collaboration{The NEXT Collaboration}

\abstract{
Double electron capture by proton-rich nuclei is a second-order nuclear process analogous to double beta decay. Despite their similarities, the decay signature is quite different, potentially providing a new channel to measure the hypothesized neutrinoless mode of these decays. The Standard-Model-allowed two-neutrino double electron capture (\ecectwonu) has been predicted for a number of isotopes, but only observed in \Kr{78}, \Ba{130} and, recently, \Xe{124}. The sensitivity to this decay establishes a benchmark for the ultimate experimental goal, namely the potential to discover also the lepton-number-violating neutrinoless version of this process, \ececnonu. Here we report on the current sensitivity of the \new detector to \Xe{124} \ecectwonu and on the extrapolation to \nexthundred. Using simulated data for the \ecectwonu signal and real data from \new operated with \Xe{124}-depleted gas as background, we define an optimal event selection that maximizes the \new sensitivity. We estimate that, for \nexthundred operated with xenon gas isotopically enriched with 1~kg of \Xe{124} and for a 5-year run, a sensitivity to the \ecectwonu half-life of $6\times 10^{22}$~y (at 90\% confidence level) or better can be reached.
}

\maketitle
       
\clearpage

\section{Introduction}
\label{sec:introduction}

Ever since the discovery of neutrino oscillations \cite{Fukuda:1998mi,Ahmad:2001an,Ahmad:2002jz}, and hence neutrino mass, there has been a sustained interest in searches for neutrinoless double beta decay processes. Such second-order weak interactions are generally considered to be the most promising way to test whether neutrinos are massive Majorana particles, identical to their anti-particles. Four related double beta processes have been proposed \cite{GomezCadenas:2011it}. The neutrinoless double $\beta^-$ emission process (\bbnonu, $(A,Z)\to (A,Z+2)+2e^-$) is possible in neutron-rich nuclei. This is, by far, the double beta process that has been explored the most from both an experimental and from a theoretical point of view. In proton-rich nuclei, up to three competing processes may be kinematically available: the neutrinoless modes of double $\beta^+$ emission, single $\beta^+$ emission plus single electron capture, and double electron capture (\ececnonu, $(A,Z)+2e^-\to (A,Z-2)$). Despite the much lower isotopic abundances of proton-rich nuclei undergoing double beta processes, \ececnonu may  provide an interesting alternative to \bbnonu searches. Particular interest in \ececnonu has been triggered by the special case in which the energy of the initial state matches precisely the energy of the (excited) final state \cite{Winter:1955zz,Bernabeu:1983yb}. If this resonance condition is met, the \ececnonu rate is expected to be increased by several orders of magnitude compared to the non-resonant (radiative) case.

\begin{figure}[b]
  \begin{center}
    \includegraphics[width=0.49\textwidth]{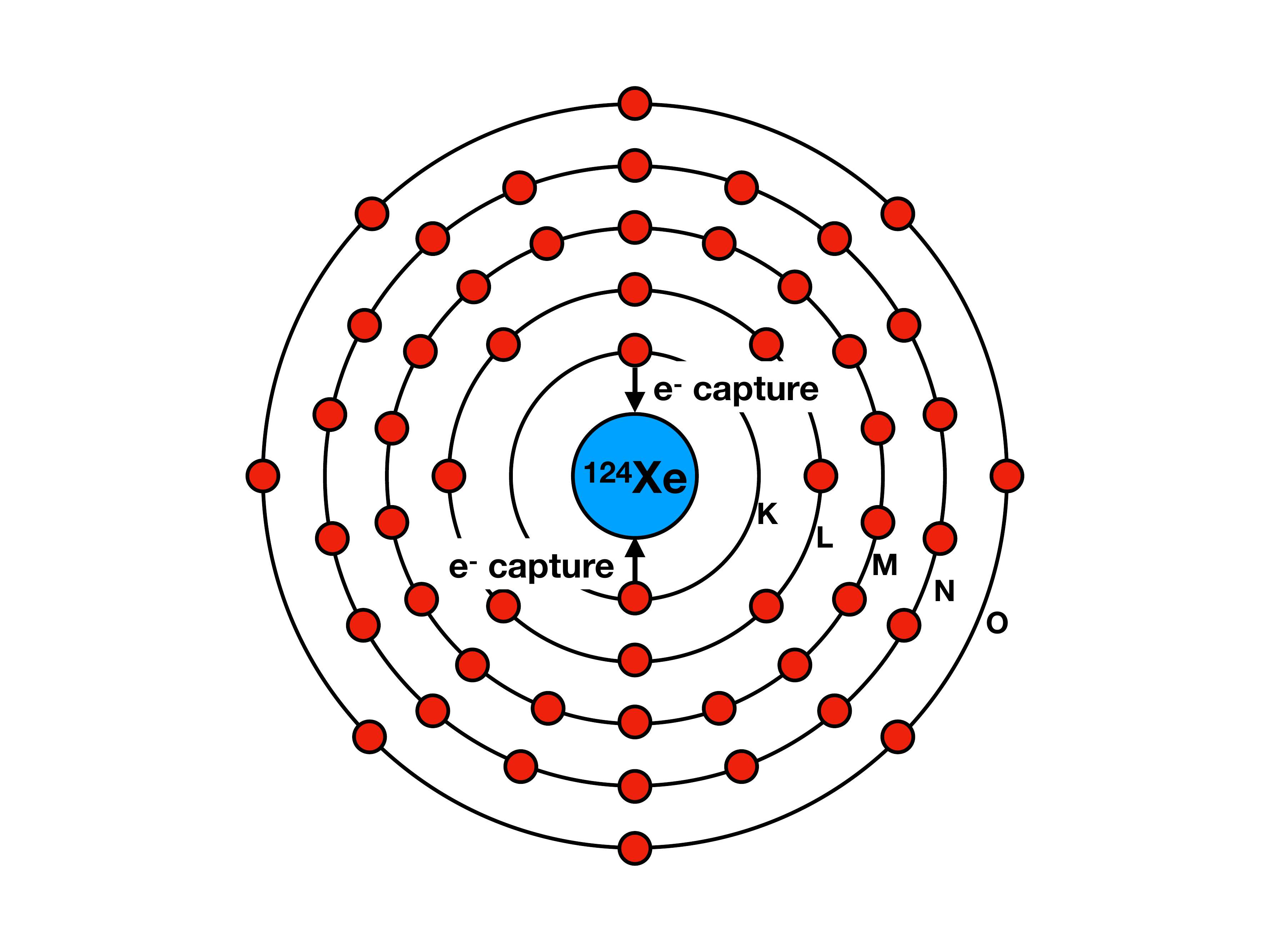} \hfill
    \includegraphics[width=0.49\textwidth]{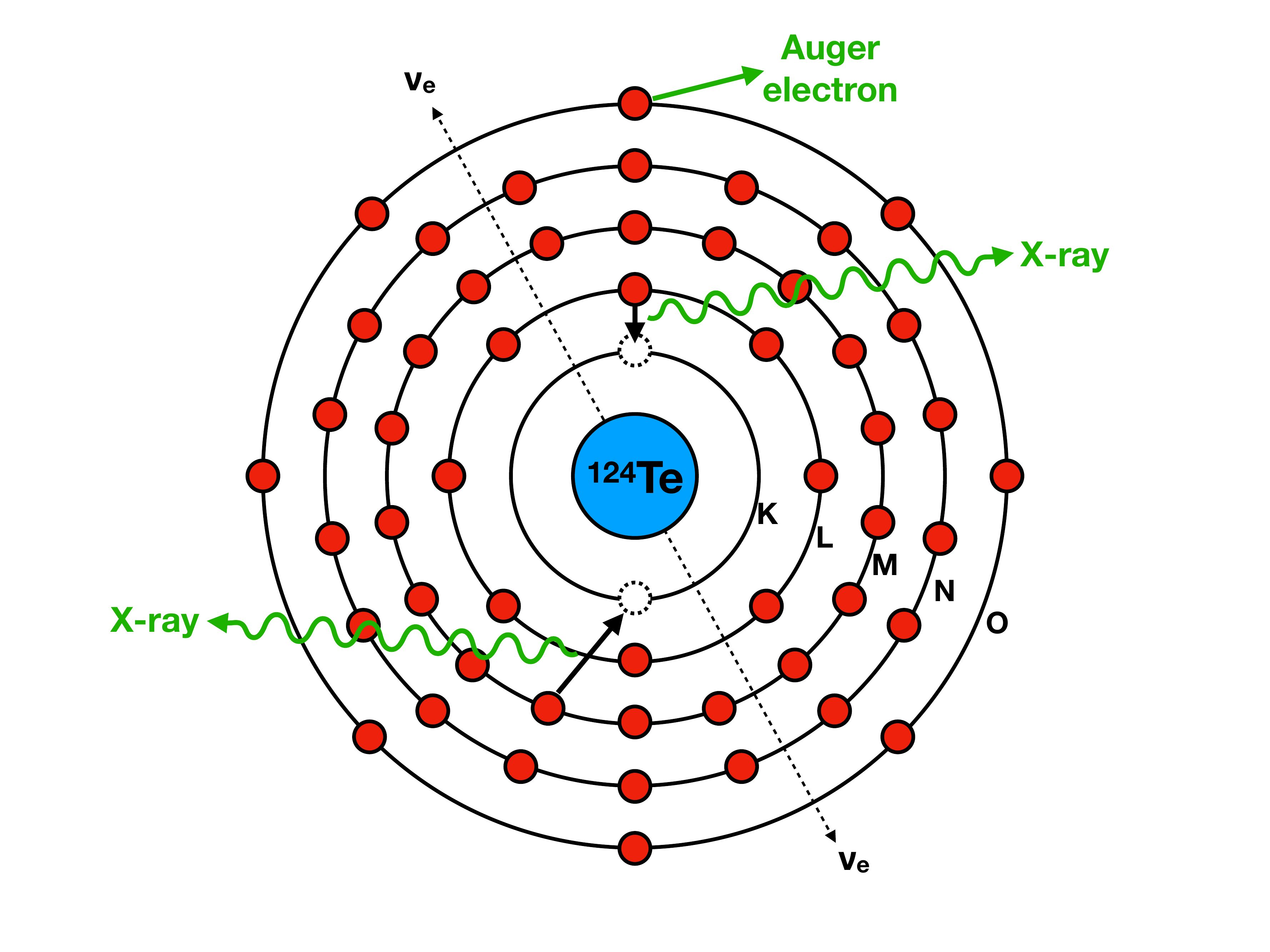}
    \caption{Schematic of the \ecectwonu process in \Xe{124}. Left: two orbital electrons from the K~shell are captured by the \Xe{124} nucleus. Right: the resulting \Te{124} atom de-excites via X-ray emission (two quanta in this example, from L$\to$K and M$\to$K transitions, respectively) or Auger electrons from the outer shells. The two electron neutrinos produced in the nuclear capture escape undetected.}
    \label{fig:ecec}
  \end{center}
\end{figure}

A fundamental step toward sensitive \ececnonu searches is the unambiguous measurement of the two-neutrino double electron capture (\ecectwonu, $(A,Z)+2e^-\to (A,Z-2)+2\nu_e$). This process is allowed in the Standard Model of particle physics, yet it is predicted to be extremely rare. The \ecectwonu predictions use similar many-body techniques and nuclear models as the ones employed for \ececnonu. Hence, a measurement of \ecectwonu rates is also relevant for interpreting \ececnonu results. In \ecectwonu, the excess energy is largely carried away by the two undetected neutrinos, with the recoil nucleus kinetic energy being too low ($\sim$ 10~eV) to be detected either. The experimental signature is hence solely given by a cascade of X-rays and Auger electrons. This cascade is a consequence of the readjustment of the electron configuration of the atom that follows the capture by the nucleus of the two orbital electrons, typically from the K~shell. The signature therefore lies in the tens of keV energy range, with all X-rays and Auger electrons produced at the same spatial location. A schematic representation of this process is shown in Figure~\ref{fig:ecec}, taking capture in \Xe{124} as an example.

\begin{table}[t]
\caption{\label{tab:xe124ecec} Summary of \Xe{124} \ecectwonu results, for electrons captured from the K~shell. For each experiment, the mass of \Xe{124} in the fiducial volume, the live time, the exposure and the half-life result are given. Lower limits are given at 90\% confidence level.}
\begin{center}
\begin{tabular}{lcccc}
  \hline
  Experiment      & \Xe{124} mass & Live time  & Exposure      & \halflife \\
  & (kg)          & (yr)       & (kg$\cdot$yr) & ($10^{22}$~yr) \\ \hline
  XENON100 \cite{Aprile:2016qsw} & 0.029 & 0.615 & 0.018 & $>0.07$ \\
  XMASS-I \cite{Abe:2018gyq} & 0.311 & 2.190 & 0.681 & $>2.1$ \\
  Gavriljuk {\it et al.} \cite{Gavriljuk:2018pez} &  0.059 & 1.760 & 0.103 & $>0.77$ \\
  XENON1T \cite{XENON:2019dti} & 1.493 & 0.487 & 0.726 & $1.8\pm 0.5$ \\ \hline
\end{tabular}
\end{center}
\end{table}

Three experimental indications of \ecectwonu exist to date. A possible evidence for \ecec has been reported in \Ba{130} from geochemical measurements of the \Xe{130} daughter, with half-life values in the (5 - 30) $\times 10^{20}$~yr range \cite{Meshik:2001ra,PUJOL20096834}. However, the half-lives of the two quoted \Ba{130} measurements are incompatible with each other. In addition, these measurements cannot separate possible contributions from the three competing double beta decay processes mentioned above, nor contributions from two-neutrino and neutrinoless modes. An indication with 4$\sigma$ significance has been reported for direct evidence of \ecectwonu in \Kr{78} with a gas proportional chamber, with a half-life of ${\halflife(\text{\Kr{78}})~=(1.9^{+1.3}_{-0.7}\pm 0.3)\times 10^{22}}$~yr~\cite{Ratkevich:2017kaz}. Recently, an observation with 4.4$\sigma$ significance has also been reported in \Xe{124} by the XENON1T Collaboration using a detector originally designed for direct dark matter searches, with \halflife(\Xe{124})=$(1.8\pm 0.5\pm 0.1)\times 10^{22}$~yr \cite{XENON:2019dti}. Details of recent \ecectwonu searches in \Xe{124} are given in Table~\ref{tab:xe124ecec}. Such searches have been performed either using large liquid xenon detectors \cite{Aprile:2016qsw,Abe:2018gyq,XENON:2019dti} with a \Xe{124} isotopic abundance close to the one of natural xenon ($9.52\times 10^{-4}$ \cite{10030100097}), or using gas proportional counters using \Xe{124}-enriched xenon \cite{Gavriljuk:2018pez}. Recent predictions from nuclear structure calculations \cite{Suhonen:2013rca,Pirinen:2015sma,Perez:2018cly} show good agreement with the \Xe{124} half-life measurement by XENON1T. In \Xe{124}, a  76.7\% fraction of all double electron captures are expected to originate from two K~shell electrons \cite{Doi:1992dm}. The majority of these double K~shell captures are expected to produce two characteristic X-rays per event.

We argue in this paper that the xenon-based high-pressure gas time projection chambers (TPCs) being developed by the NEXT Collaboration for \Xe{136} \bbnonu searches are ideally suited to also search for double electron capture in \Xe{124}, provided that a gas with sufficient \Xe{124} fraction is used. From simulated signal data and background data taken with the \new detector we optimize the data selection cuts that provide the best sensitivity to the \ecectwonu decay. These data are finally used to extrapolate the sensitivity to the \nexthundred detector.

The paper  is organized as follows. Section~\ref{sec:next} describes the NEXT experimental program, with special emphasis in the NEXT-White detector used in this feasibility study. Section~\ref{sec:data} describes the background data and signal simulation samples employed.  The event selection and the  sensitivity projections  are presented in Sections~\ref{sec:selection} and \ref{sec:sensitivity}, respectively. We conclude in Section~\ref{sec:conclusions}.

\section{The NEXT experimental program}
\label{sec:next}

The NEXT detectors rely on the technology of high-pressure (10--15~bar) xenon gas time projection chambers (TPCs) with electroluminescent (EL) amplification and optical readout. The gas TPC provides a homogeneous and low-density detector active volume. Electroluminescence provides a nearly noiseless amplification of the ionization electrons reaching the TPC anode, key for excellent energy resolution capabilities. The optical signals induced by xenon primary (\Sone) and electroluminescent (\Stwo) scintillation are detected by readout planes at opposite ends of the detector's cylindrical volume, behind transparent cathode and anode planes, capable of providing a full 3D image of the events. The so-called tracking plane is composed of a large number of silicon photomultipliers (SiPMs) placed on a 2D lattice, and is located a few mm away from the anode and the EL region. The tracking plane detects the forward-going \Stwo light, providing an unambiguous 2D image for each TPC time slice. The so-called energy plane is composed of photomultiplier tubes (PMTs) and is located behind the cathode, away from the EL region. The energy plane detects the nearly uniform backward-going \Stwo light, providing an accurate energy measurement. The energy plane also detects the prompt \Sone light produced in the drift volume, enabling an accurate event $t_0$ determination.

\begin{figure}[t]
  \begin{center}
    \includegraphics[trim= 4.cm 4.cm 11.cm 0., clip, width=0.45\textwidth]{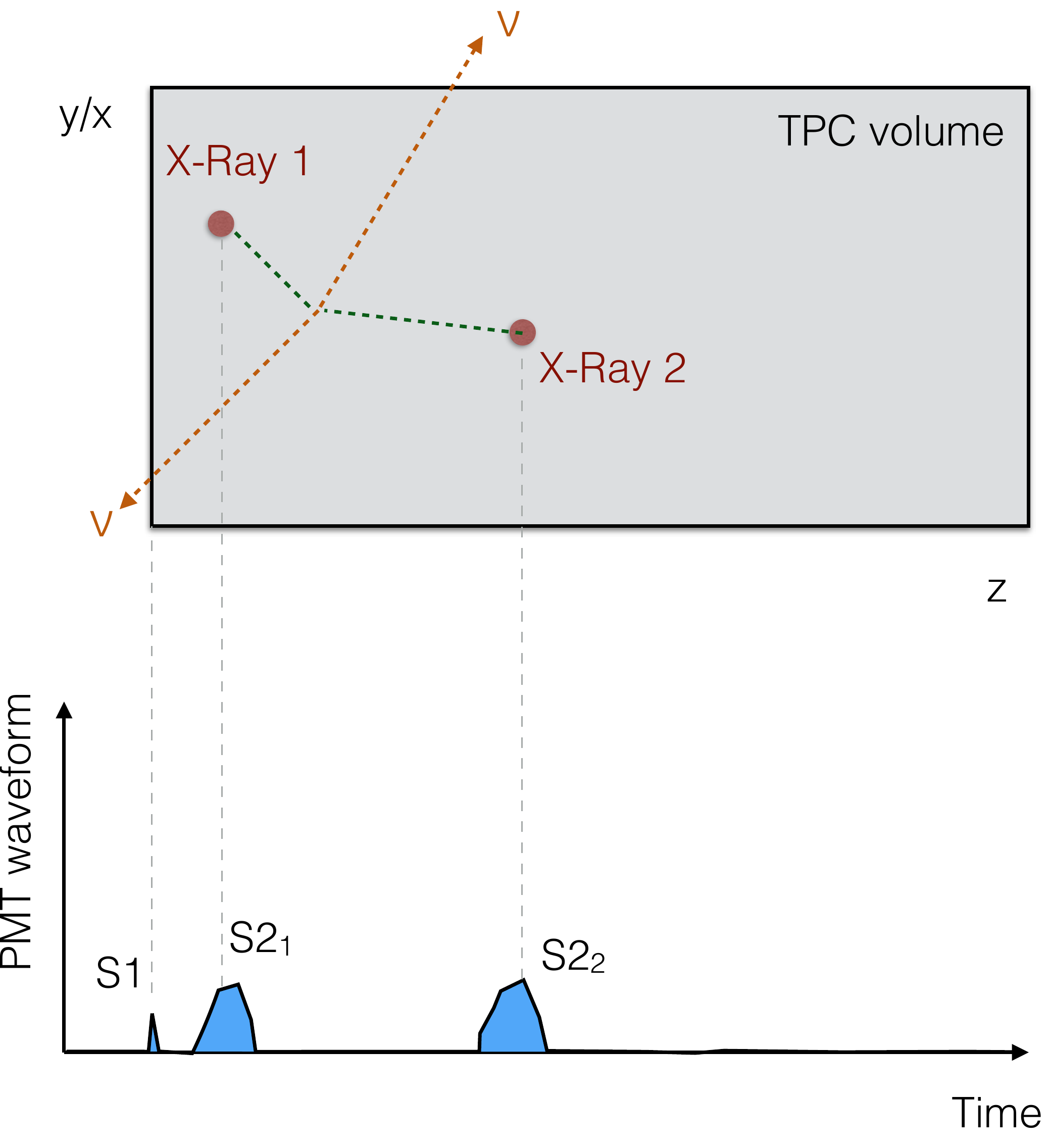} \hfill
    \includegraphics[trim= 4.cm 4.cm 11.cm 0., clip, width=0.45\textwidth]{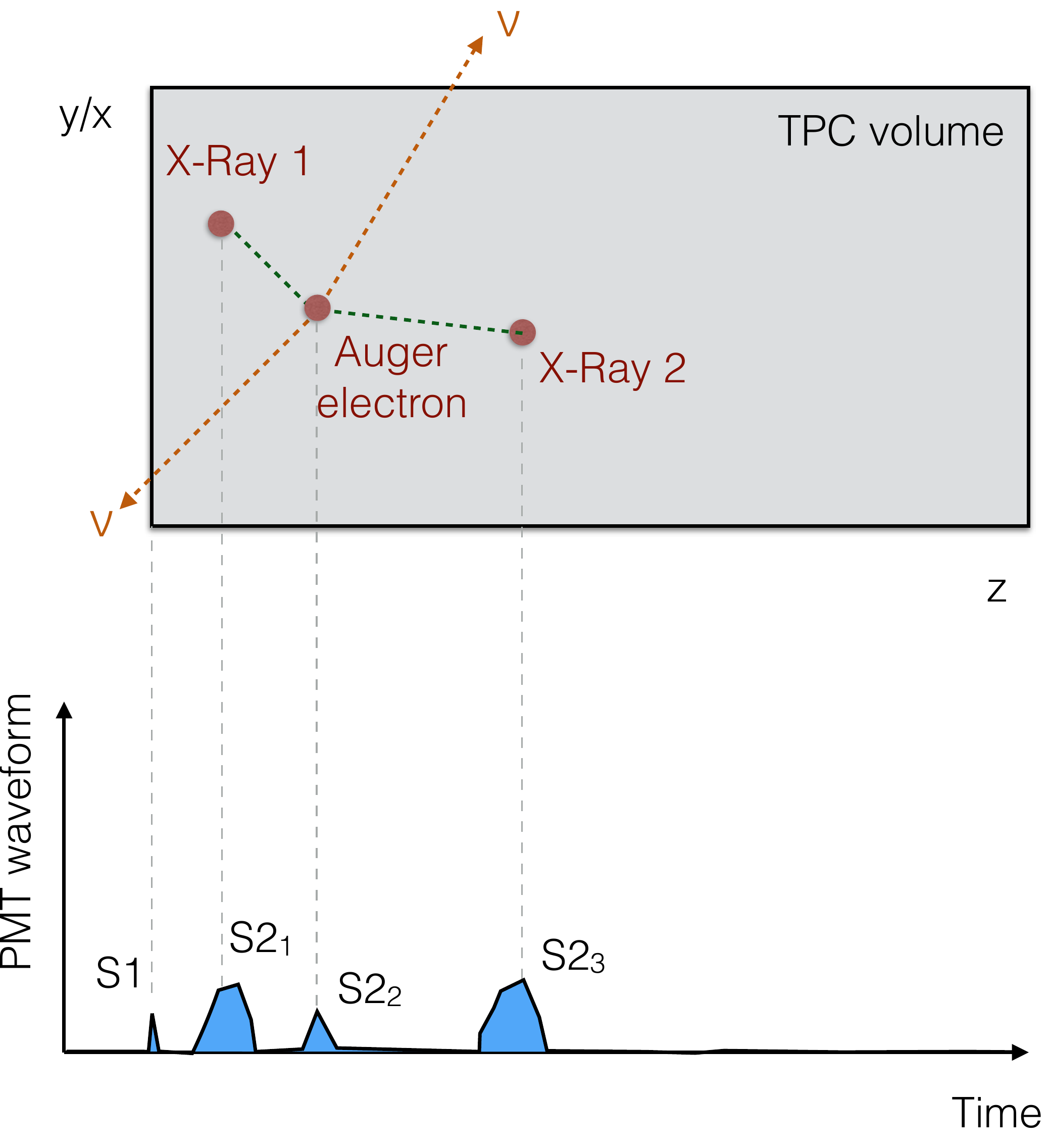}
    \caption{Schematic of \ecectwonu signature in NEXT. Two golden samples are considered, with two (left panel) and three (right panel) separate energy depositions per event, respectively.}
    \label{fig:ecec_next}
  \end{center}
\end{figure}

The NEXT detection concept applied to \ecectwonu signals is depicted schematically in Figure~\ref{fig:ecec_next}. Each X-ray interaction in the active volume may give rise to a separate \Stwo signal, some millimeters or centimeters away from the event vertex. For example, tellurium $K_{\alpha}$ X-rays from L$\to$K transitions produced by \Xe{124} double electron capture, with an energy of 27.5~keV, have a linear attenuation coefficient of 1.6~cm in xenon at 10~bar. A weaker \Stwo signal from Auger electrons at the event vertex may also be reconstructed. The \Stwo signals have a pulse shape characteristic of point-like energy deposits, with a time width primarily affected by diffusion effects along ionization electron drift. Regardless of the number of \Stwo signals, a single (narrower) \Sone signal characteristic of the full event energy deposition is also present in the event PMT waveforms.

The strengths of the NEXT approach are threefold. First, NEXT TPCs feature a better energy resolution in the energy region of interest compared to the liquid xenon scintillators or gas proportional counters listed in Table~\ref{tab:xe124ecec}. Second, NEXT provides full 3D position reconstruction capabilities to suppress external background events. This is also the case of liquid scintillators (which also provide xenon self-shielding), but only partially the case for the gas proportional counter of reference \cite{Gavriljuk:2018pez}. Third, the low density of the detector and its 3D imaging capabilities allow NEXT to spatially separate the X-ray conversions or Auger electron deposits for a significant fraction of all \ecectwonu events.  This is not possible in the higher-density liquid xenon, and only partially possible (that is, along the drift direction) in a gas proportional counter.

The first phase of the NEXT experimental program started in 2009 with the construction, commissioning and operation of two EL prototypes, NEXT-DEMO \cite{Alvarez:2013gxa} and NEXT-DBDM \cite{Alvarez:2012yxw}, with xenon active masses of  about 1~kg. These prototypes demonstrated the robustness of the technology, its excellent energy resolution and its unique topological signature. The NEXT-White demonstrator \cite{Monrabal:2018xlr}, deploying 4~kg of xenon in its active volume, implements the second phase of the program. This detector has been operating underground at the Laboratorio Subterr\'aneo de Canfranc (LSC, Spain) since 2016. NEXT-White is also the first radiopure detector in the NEXT series. Its main goals are a detailed assessment of the backgrounds for \Xe{136} double beta decay searches, the measurement of the \Xe{136} \bbtwonu half-life and the characterization of the detector performance at energies close to the \Xe{136} energy region of interest (about 2.5~MeV). The \nexthundred detector, currently under construction, constitutes the third phase of the program. With a \Xe{136} active mass approaching 100~kg, \nexthundred will perform the first  sensitive \bbnonu search in xenon gas \cite{Martin-Albo:2015rhw}. The fourth phase of the program contemplates tonne-scale xenon gas detectors. Two R\&D lines are being pursued in parallel. The NEXT technology can be scaled up to \bbnonu source masses in the tonne scale introducing several technological advancements already available \cite{Adams:2020cye,Giuliani:2019uno}. The NEXT Collaboration is also pursuing a more radical approach to a tonne-scale experiment based on the efficient detection of the Ba$^{++}$ ion produced in the \bbnonu decay of \Xe{136} using single-molecule fluorescence imaging (SMFI) \cite{Giuliani:2019uno,Nygren:2015xxi,Jones:2016qiq,McDonald:2017izm,Thapa:2019zjk,Rivilla:2019vzd}.

\section{Data samples}
\label{sec:data}

For a given \Xe{124} mass in the detector, a reliable evaluation of the experimental sensitivity to \Xe{124} \ecectwonu relies on two factors: a good description of the backgrounds and a realistic estimate of the efficiency in reconstructing and selecting \ecectwonu signal events. The lower the backgrond rate and the higher the signal efficiency, the better the \Xe{124} \ecectwonu sensitivity and the prospects for its observation. For this analysis, we combine a background dataset from the \new detector adding up to 5 months of data-taking (Section~\ref{subsec:data_background}) with a custom Monte-Carlo (MC) simulation of the \ecec signal (Section~\ref{subsec:data_signal}).


\subsection{Background data sample}
\label{subsec:data_background}

\begin{figure}[t]
  \begin{center}
    \includegraphics[width=0.7\textwidth]{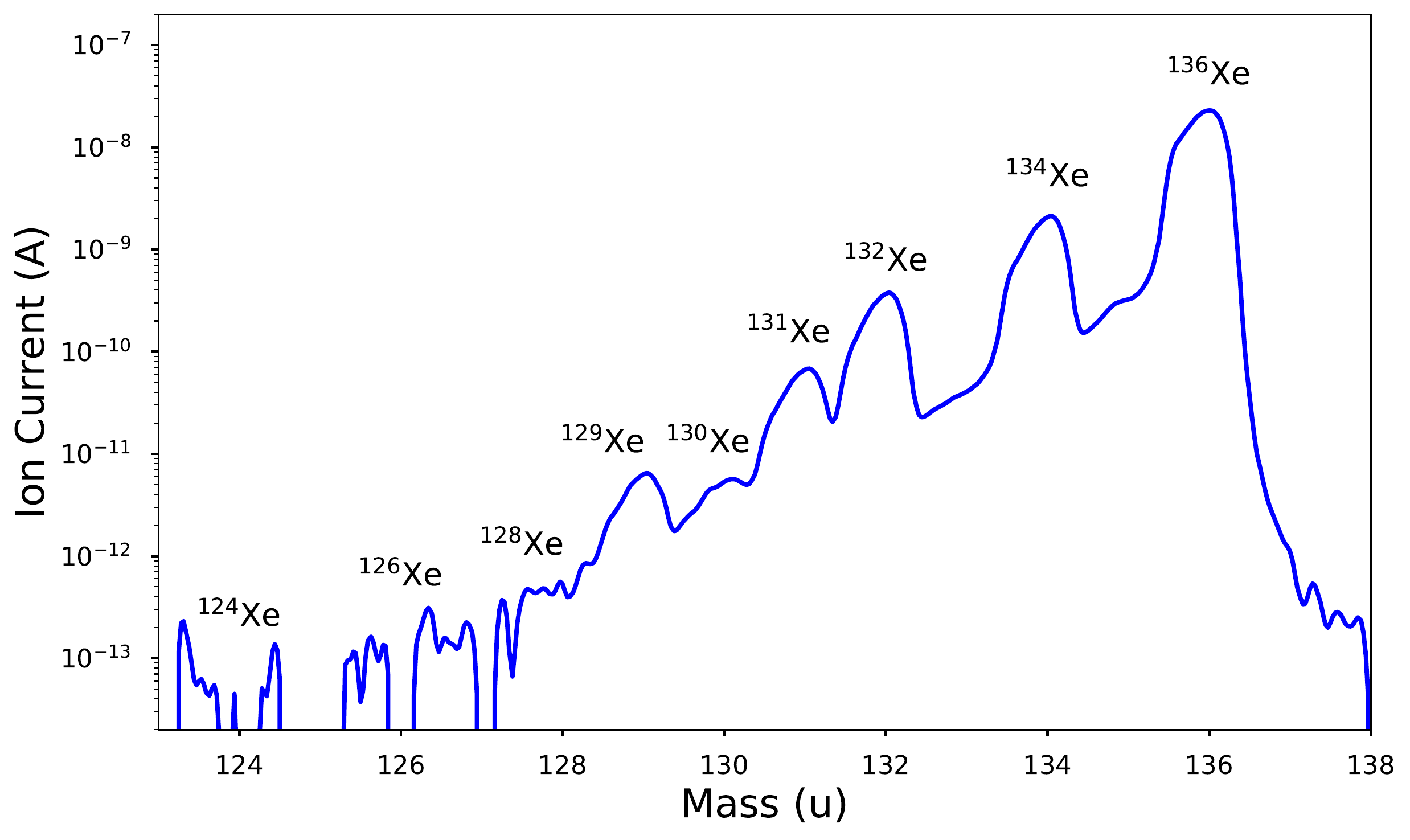}
    \caption{Isotopic composition of NEXT \Xe{136}-enriched gas.}
    \label{fig:xenon}
  \end{center}
\end{figure}

The background sample used in this analysis was taken in 2019 between February 25th and July 10th and between September 13th and November 6th, as part of Run~Va and Run~Vb of the \new detector, which was filled with \Xe{136}-enriched gas.
The amount of \Xe{124} present in the source is negligible, as shown in Figure~\ref{fig:xenon}, which makes this run configuration perfect for background characterization.
The total accumulated exposure in this period is 125.9 days.

The events considered in the analysis are those taken by the low-energy trigger of the detector, which is set to record events between approximately 8~keV and 80~keV.
This background dataset constitutes a total of $2.2\times 10^8$ low-energy triggers.
Detector conditions were stable during the whole data-taking period.
The pressure was 10.13 bar, with voltages of 30 kV and 7.7 kV in the cathode and gate.
The drift velocity of the electrons in the gas \cite{Simon:2018vep} remained fairly constant at $\sim$0.91 mm/$\mu$s.
The electron lifetime during this period of data-taking varied between 5 and 9.5 ms, many times greater than the maximum drift time.
The smooth variations in the detector conditions are accounted for using \Kr{83\mathrm{m}} calibration data, which allows for a continuous monitoring of the electron lifetime and other detector properties \cite{Martinez-Lema:2018ibw}.

Due to the low-energy nature of these events, the data are processed to obtain a full 3D point-like reconstruction for each separate energy deposit in the event.
The waveforms from the PMTs are summed to identify \Sone and \Stwo signals.
The integral of each \Stwo signal is used to determine the deposited energy in each site.
The waveforms of the SiPM sensors are combined to perform a 2D reconstruction of each \Stwo signal separately, which is in turn aggregated with the drift time to have one point-like 3D reconstruction per \Stwo signal.
The energy of each \Stwo signal is finally corrected by lifetime and geometry effects and converted to keV on a run-by-run basis. For more details on NEXT-White low-energy data processing and calibration procedures, see reference \cite{Martinez-Lema:2018ibw}.


\subsection{Signal simulation sample}
\label{subsec:data_signal}

The only particles emitted in the \ecec process are neutrinos, which are undetectable.
Hence, the simulation of the decay is focused on the daughter nucleus, which for \Xe{124} is \Te{124}.
Although electrons of any shell can be captured, K~shell electrons have the highest probability.
Thus, the aim is to study \Te{124} atoms with a double K~shell vacancy, which relax by emitting X-rays or Auger electrons.
The simulation is based on the atomic relaxation package of Geant4 v10.04 \cite{Agostinelli:2002hh}. Geant4 uses the Livermore Evaluation Atomic Data Library (EADL) \cite{Perkins:236347}, that contains data to describe the relaxation of atoms back to neutrality after they are ionized.
Since the package does not provide a mechanism to generate \ecec events, we approximate our simulation by generating two independent \Te{124} atoms with a single K~shell vacancy and at the same spatial location per event, as done in reference \cite{Abe:2018gyq}.

Albeit the energy of two single K~shell vacancies (63.63~keV \cite{RevModPhys.39.125})
is not the same of that of a double vacancy (64.46~keV \cite{PhysRevC.86.044313}), we consider the difference, 0.8~keV, negligible.

\begin{figure}[t]
  \begin{center}
    \includegraphics[width=0.49\textwidth]{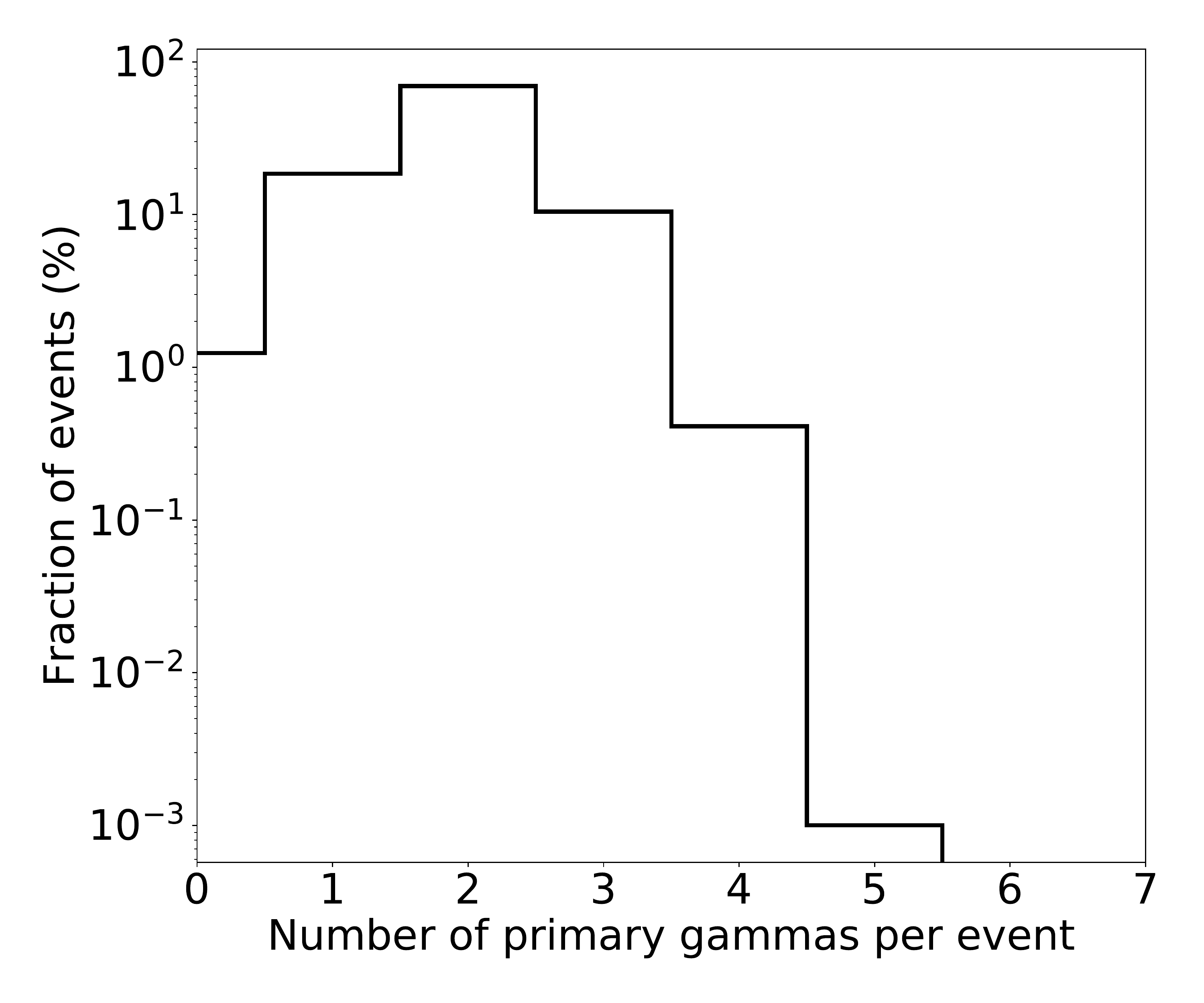} \hfill
    \includegraphics[width=0.49\textwidth]{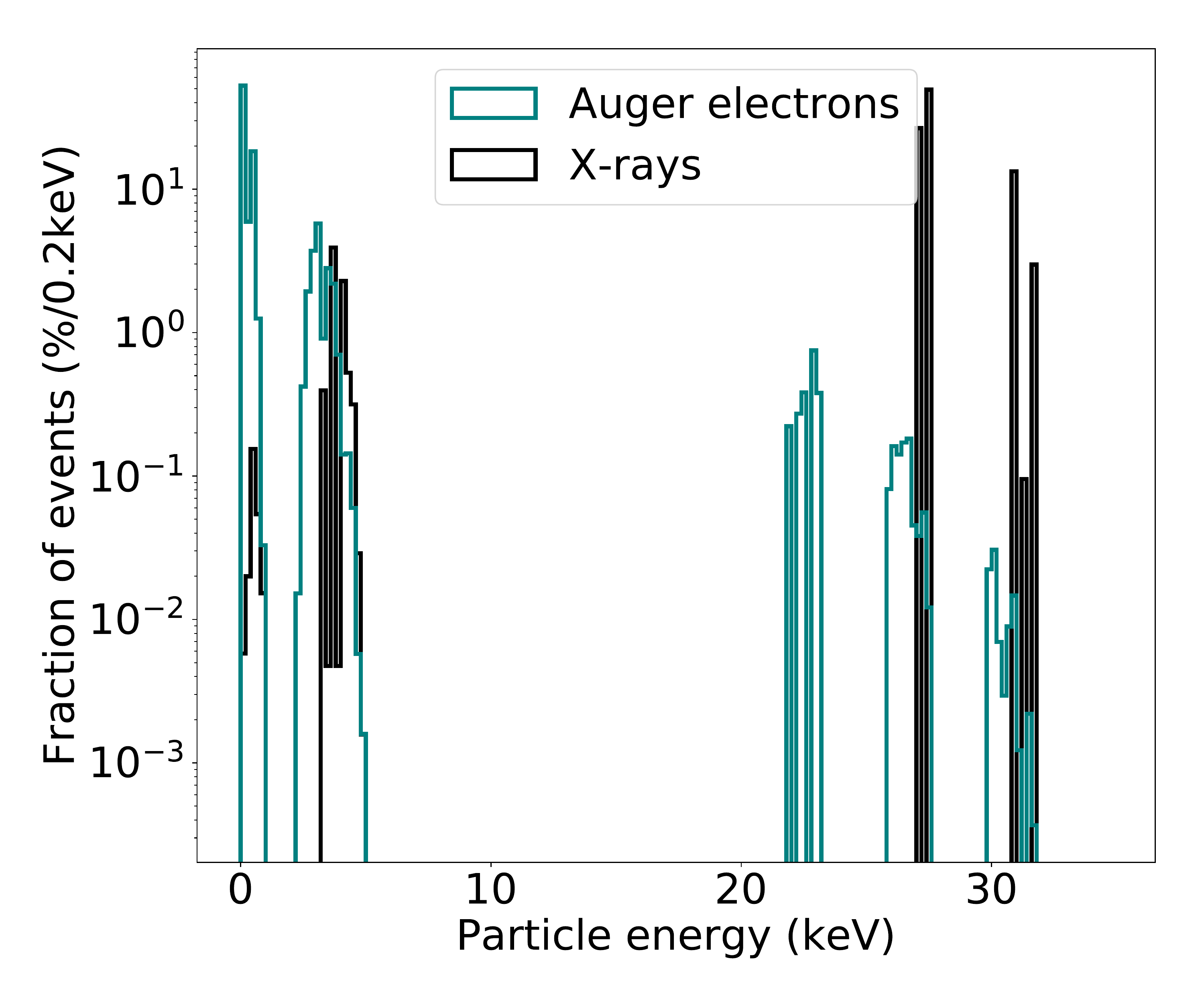}
    \caption{Monte-Carlo truth distributions of \Xe{124} \ecec signal events. Left: number of primary gammas (X-rays) per simulated event. Right: distribution of simulated X-ray energies.}
    \label{fig:mctruth}
  \end{center}
\end{figure}

Each atom in the simulation de-excites independently according to the X-ray and Auger emission probabilities.
The observed emission probabilities, derived from the number of primary gammas in the simulation shown in the left panel of Figure~\ref{fig:mctruth} are consistent with the fluorescence yield of \Te{124} \cite{Firestone.1997}.
The energy of the emitted gammas matches also the atomic energy levels of the \Te{124} atom, as shown on the right panel of Figure~\ref{fig:mctruth}.

The interactions of electrons and gammas in the volume of the detector are simulated according to the \textit{G4EmStandardPhysics\_option4} package of Geant4.
Then, we simulate the electron drift (with realistic electron diffusion and negligible electron attachment), the light emission using parametrized models of the detector and the electronics response (including noise) to finally produce a set of sensor waveforms, equivalent to that of the real detector.
These waveforms are then processed identically as detailed in Section~\ref{subsec:data_background}.

The full signal dataset is composed of $10^5$ events distributed homogeneously over the active volume of the \new detector.

\section{Event selection}
\label{sec:selection}

Signal and background events produce different signatures in the detector.
At low energies ($<$100~keV), the events in the chamber are primarily small-angle Compton scatterings of background gamma-rays, photoelectric interactions of low-energy gamma-rays, isolated xenon X-rays and, mostly, decays of metastable \Kr{83\mathrm{m}} atoms introduced in the active volume as a calibration source~\cite{Martinez-Lema:2018ibw}.
These events are characterized by having a single-site interaction and thus producing one \Sone and one \Stwo signals.
On the other hand, the vast majority of \ecectwonu events consist of at least two interactions, regardless of whether they come from X-rays or Auger electrons.  This can be inferred for example from the left panel of Figure~\ref{fig:mctruth}, showing that approximately 99\% of \ecectwonu events are expected to produce at least one  X-ray. Hence, the event topology provides a key feature to discriminate background events from signal.

Notwithstanding, there are two cases in which this feature is not as useful.
First, pile-up of different events in the DAQ window could in principle mimic the signature of a \ecec event, if only one \Sone signal is reconstructed.
However, the energies of the individual \Stwo signals rarely match the ones expected from our signal events.
Second, the two X-rays or Auger electrons from a \ecec event can interact very close to the originating vertex, meaning that both signals merge into a single \Stwo, spoiling the topological signature of the process.
A fraction of these events could be recovered by improving and customizing the reconstruction algorithms for this particular case, see discussion in Section~\ref{subsec:sensitivity_improved}.

In order to optimize the signal and minimize the background, we perform a number of selections in the data.
These selections are based both on topology and energy considerations.
First, we reduce the data size by selecting a broad energy window around the region of interest (ROI), which eliminates most of the events coming from the \Kr{83\mathrm{m}} source, see Section~\ref{subsec:selection_krveto}.
Second, we apply data quality cuts in Section~\ref{subsec:selection_quality}.
In Section~\ref{subsec:selection_fiducial} we discuss the fiducial cuts.
We then select the optimal energy ROI for the search, see Section~\ref{subsec:selection_energy}. We finally apply the multi-site event selection in Section~\ref{subsec:selection_multisite}. Each selection is done separately and based on the optimization of the following figure of merit:
\begin{equation}
  \textrm{FOM} = \frac{\varepsilon_{\mathrm{sig}}}{\sqrt{\varepsilon_{\mathrm{bkg}}}},
  \label{eq:fom}
\end{equation}
where $\varepsilon$ is the relative efficiency of the cut defined as
\begin{equation}
  \varepsilon = \frac{\textrm{\# events after the cut}}{\textrm{\# events before the cut}}.
  \label{eq:efficiency}
\end{equation}

A data reduction summary for signal and background events as a function of the various sequential cuts is given in Table~\ref{tab:selection} and discussed in Section~\ref{subsec:selection_summary}.

\begin{table}[t]
  \caption{\label{tab:selection}Event selection summary. Cumulative signal and background efficiency as a function of the various event selection criteria.}
\begin{center}
\begin{tabular}{lcc}
  \hline
  Processing/selection step  & Fraction of          & Fraction of    \\
                             & background data (\%) & signal MC (\%) \\ \hline
  DAQ triggers or MC simulated events & 100 & 100 \\
  Event reconstruction      &     83.564 $\pm$ 0.003                             &   89.10 $\pm$ 0.10 \\
  \Kr{83\mathrm{m}} veto    &    (40.53  $\pm$ 0.04 )$\times \mathrm{10^{-2}}$   &   83.50 $\pm$ 0.12 \\
  Data quality              &    (13.99  $\pm$ 0.02 )$\times \mathrm{10^{-2}}$   &   35.76 $\pm$ 0.15 \\
  Fiducial                  &    (4.06   $\pm$ 0.04 )$\times \mathrm{10^{-3}}$   &   25.42 $\pm$ 0.14 \\
  Event energy              &    (5.28   $\pm$ 0.15 )$\times \mathrm{10^{-4}}$   &   24.39 $\pm$ 0.13 \\
  Multi-site                &    (1.15   $\pm$ 0.07 )$\times \mathrm{10^{-4}}$   &   22.90 $\pm$ 0.13 \\ \hline
\end{tabular}
\end{center}
\end{table}


\subsection{\Kr{83\mathrm{m}} veto requirement}
\label{subsec:selection_krveto}

The \Kr{83\mathrm{m}} calibration source used in the experiment produces 41.5~keV single-site events that represent about 82\% of the reconstructed low-energy events.
The energy resolution of the NEXT-White detector at 41.5~keV is approximately 1.8~keV~FWHM~\cite{Martinez-Lema:2018ibw}. A $1/\sqrt{E}$ extrapolation, demonstrated in \cite{Renner:2019pfe}, yields an expected energy resolution at 63.63~keV of 2.2 keV. Thus, we can reduce the \Kr{83m} background to the minimum by requiring the energy of the entire event to be in a broad window above the \Kr{83\mathrm{m}} peak.

\begin{figure}[t]
  \begin{center}
    \includegraphics[width=0.5\textwidth]{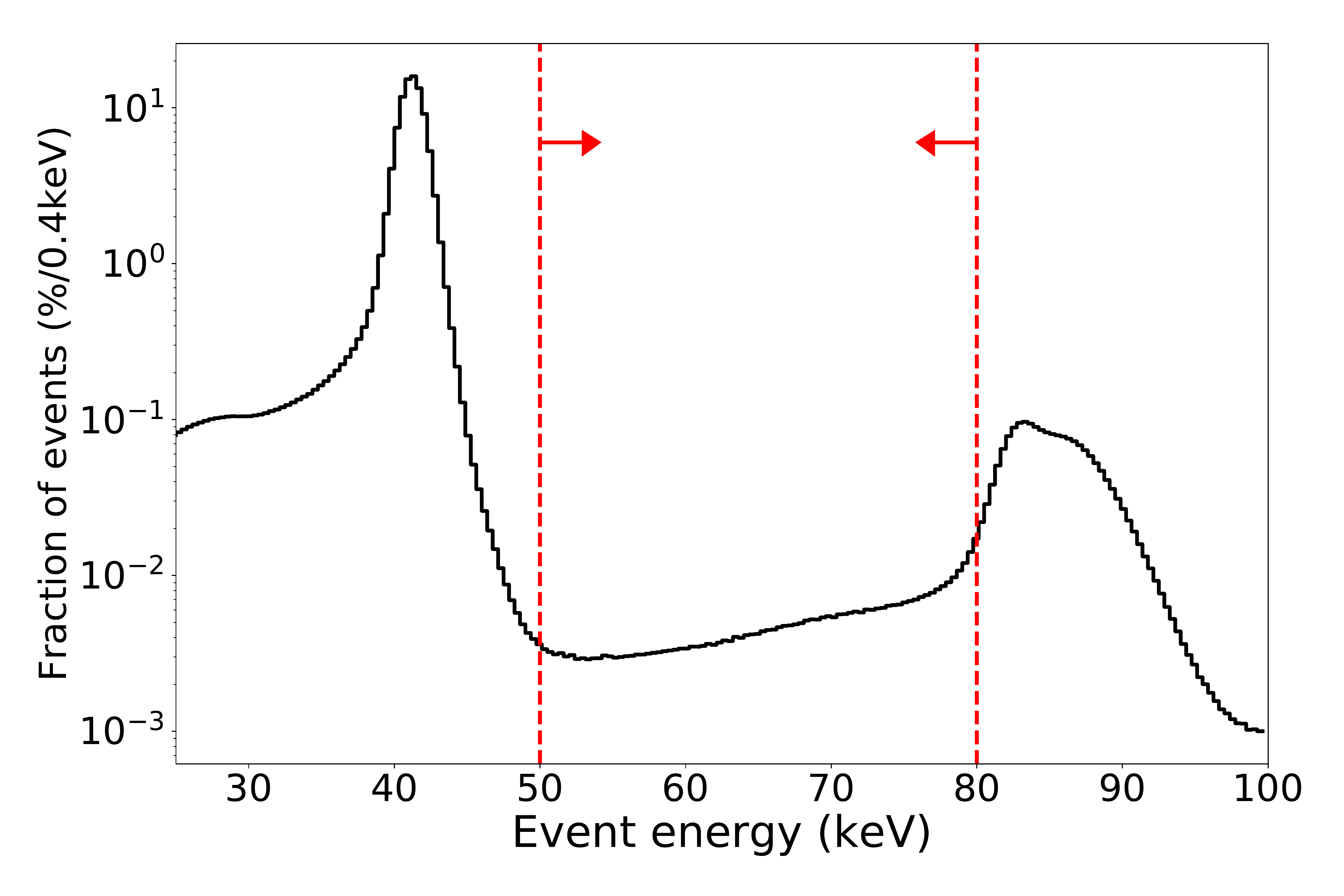}
    \caption{Event energy spectrum of fully reconstructed low-energy triggers for the full background dataset taken with the \new detector. Red lines indicate the broad energy window for the \Kr{83\mathrm{m}} veto.}
    \label{fig:bkg_spectrum}
  \end{center}
\end{figure}

As shown in Figure~\ref{fig:bkg_spectrum}, the energy spectrum is dominated by the \Kr{83\mathrm{m}} peak at 41.5~keV and, to a lesser extent, by the coincidence of two \Kr{83\mathrm{m}} events in the same waveform at 83~keV.
We remove these data from our sample by selecting the events with a total energy between 50~keV~$< E_{\mathrm{evt}} < 80$~keV.
The combined requirement of successful event reconstruction plus \Kr{83\mathrm{m}} veto keeps 0.4\% of the background data, hence reducing the size of the data sample by a factor of 200, while keeping 83.5\% of the simulated \ecectwonu signal events.

\subsection{Data quality selections}
\label{subsec:selection_quality}

Data quality criteria are imposed based on the multiplicity of reconstructed \Sone and \Stwo signals per event. First, we remove all events with no or multiple \Sone signals.
Events with no \Sone signals are not reconstructed along the drift coordinate\footnote{The drift time can also be estimated from the width of the S2 signal. This method is less accurate, but it might be required in larger detectors where the \Sone detection efficiency is smaller.}. Therefore, they cannot be selected according to fiducial volume criteria nor their energy corrected for electron lifetime effects along drift.
Events with more than one \Sone signal on the other hand introduce an ambiguity in the drift coordinate determination, resulting in the same limitations as events with no \Sone signals. Additionally, such events may originate from event pile-up.
The single \Sone condition keeps 45.3\% of the background data and  100\% of the simulated signal.

\begin{figure}[t]
  \begin{center}
    \includegraphics[width=0.5\textwidth]{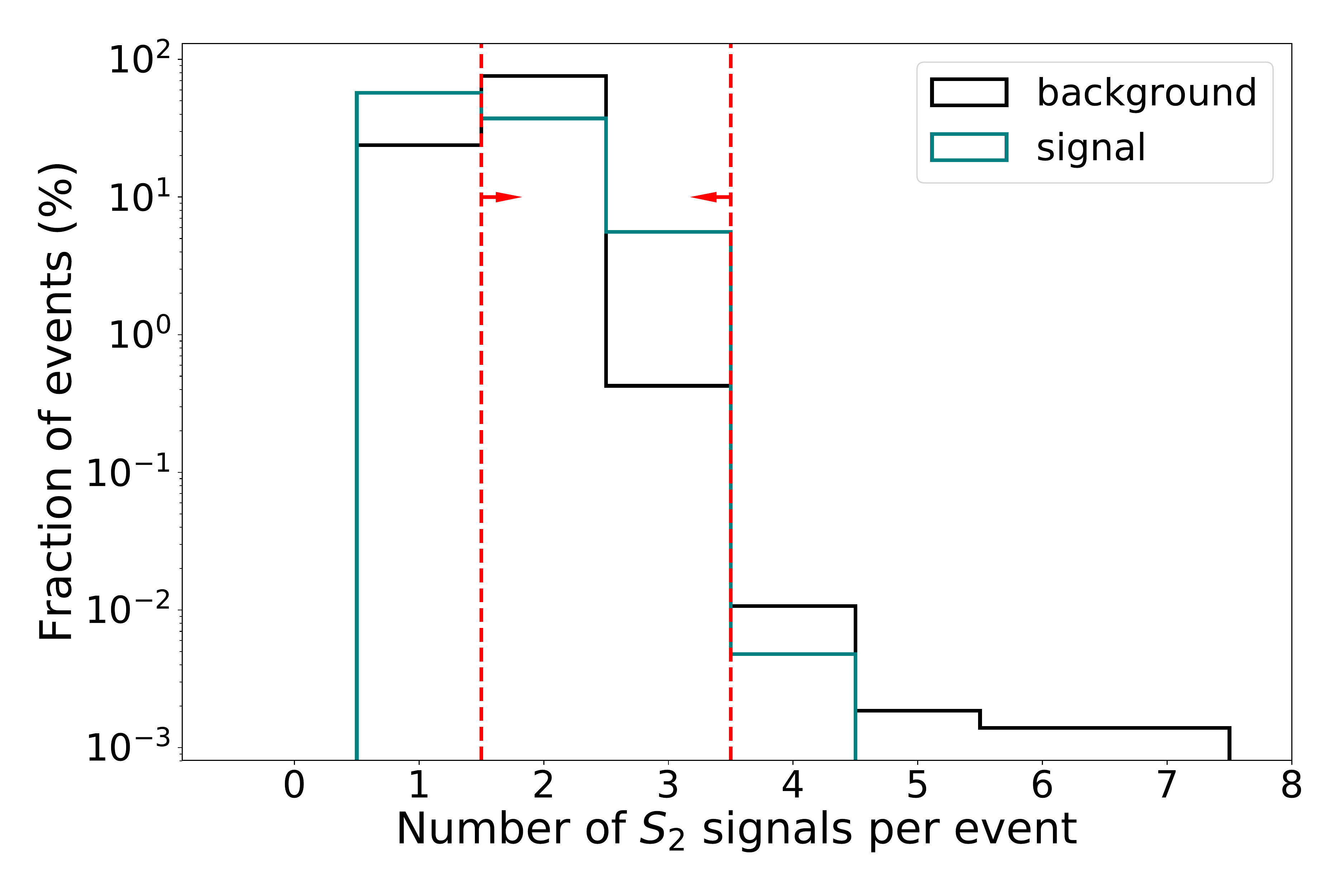}
    \caption{\Stwo multiplicity per event for signal and background events after the \Kr{83\mathrm{m}} veto. Red lines indicate the limits of the range of accepted values.}
    \label{fig:s2multiplicity}
  \end{center}
\end{figure}

Second, we require each event to have either two or three \Stwo signals.
This condition is based on the topology of \ecectwonu events, which should produce two or three separate energy depositions in most cases, see Figures~\ref{fig:ecec_next} and \ref{fig:mctruth}.
Thus, this requirement selects events in which at least two of them are spatially distinguishable along the drift coordinate. This cut keeps 42.8\% of the signal events and 76\% of the background events. Despite losing a significant fraction of signal events, for this study we assume that energy depositions that overlap in Z are not separable as doing so would require dedicated reconstruction algorithms as discussed in Section \ref{subsec:sensitivity_improved}.
As shown in Figure~\ref{fig:s2multiplicity}, the probability of reconstructing three \Stwo signals per event after the \Kr{83\mathrm{m}} veto is higher in signal than in background, while the opposite is true for two-\Stwo events.

\subsection{Fiducial selection}
\label{subsec:selection_fiducial}

Signal events are homogeneously distributed over the entire active volume.
On the other hand, background events tend to be reconstructed on the borders of the detector. Thus, we define a fiducial volume of the detector that maximizes the sensitivity to our search.
An event is considered to pass the fiducial cut if and only if all of the \Stwo signals satisfy the condition. Three cut variables are separately optimized to define the fiducial selection: the maximum radial position $R_{\mathrm{max}}$, the minimum drift distance $Z_{\mathrm{min}}$ and the maximum drift distance $Z_{\mathrm{max}}$ of all \Stwo signals in the event. In order to obtain the optimal selection we evaluate the figure of merit in Eq.~(\ref{eq:fom}) for each variable independently within a sensible range of values for the cut variables.

\begin{figure}[t]
  \begin{center}
    \includegraphics[width=0.49\textwidth]{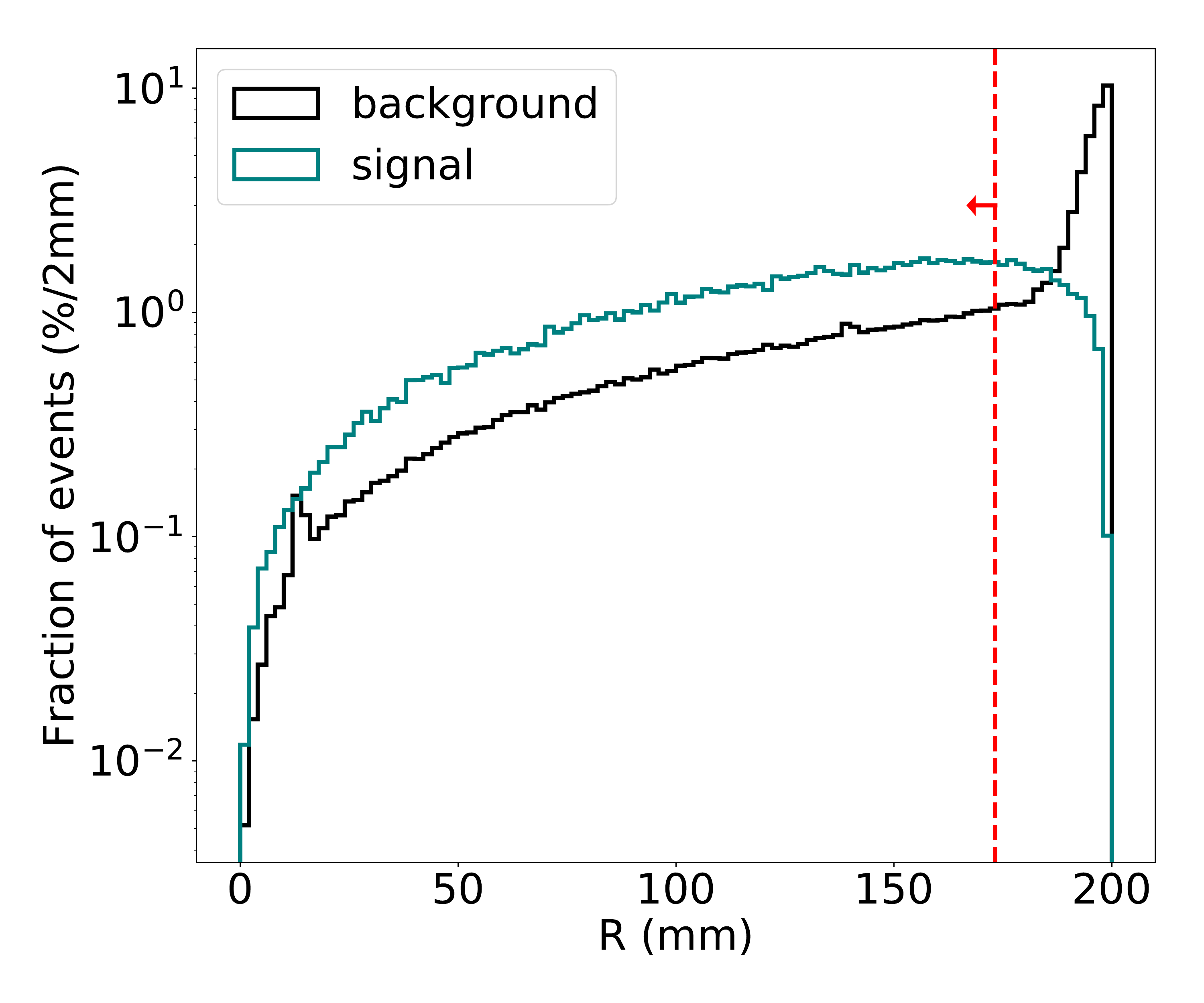} \hfill
    \includegraphics[width=0.49\textwidth]{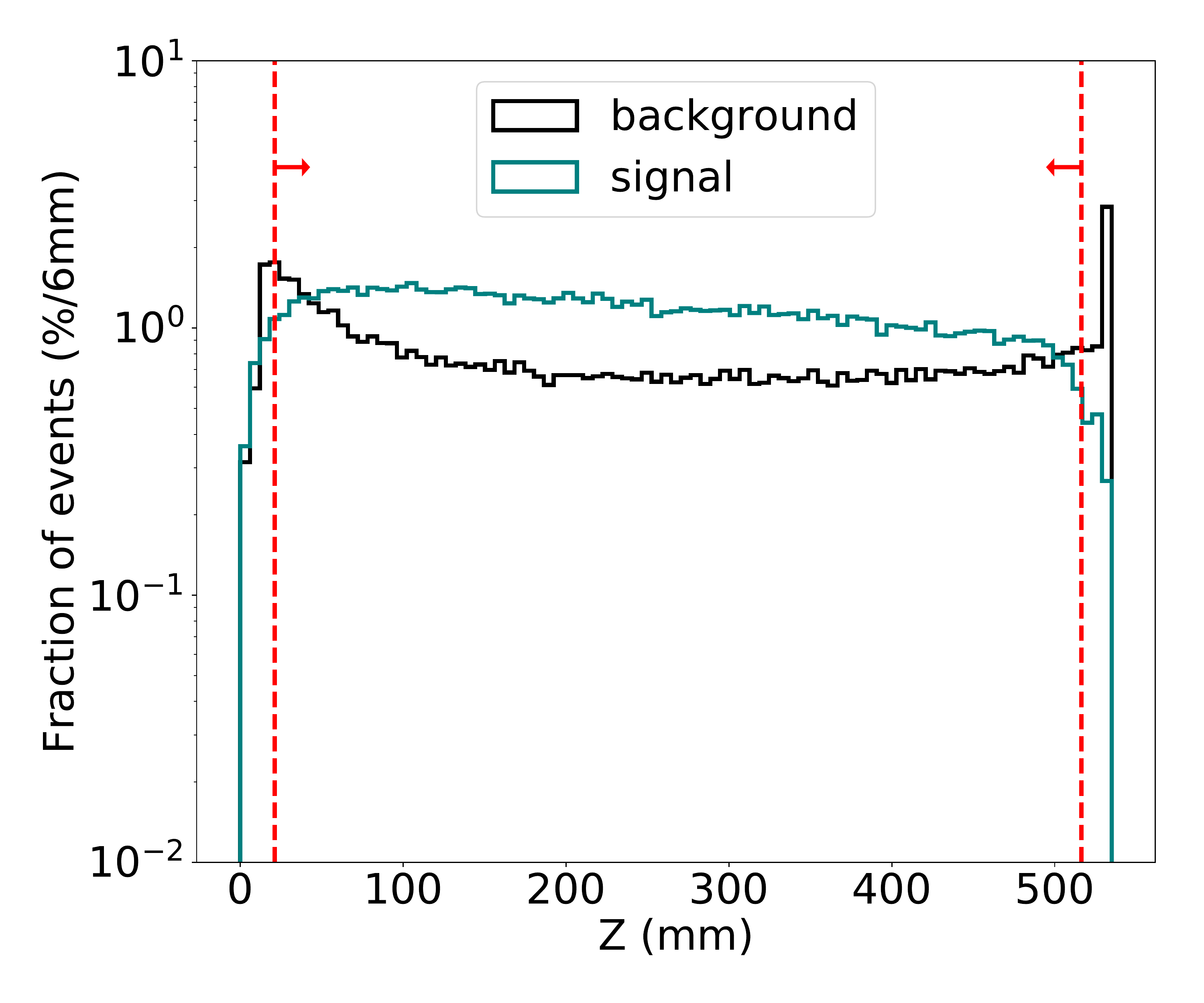}
    \caption{Event selection based on spatial information. One entry per reconstructed \Stwo signal is shown. Left: signal and background distributions along the radial detector coordinate, with optimal $R_{\mathrm{max}}$ cut position (red line). Right: signal and background distributions along the drift coordinate and after the fiducial radius selection, with optimal $Z_{\mathrm{min}}$ and $Z_{\mathrm{max}}$ cut positions (red lines).}
    \label{fig:selection_fiducial}
  \end{center}
\end{figure}

For the radial coordinate we only set an upper limit. The optimal value is found to be $R_{\mathrm{max}}=183.5$~mm, to be compared with the \new active volume radius of $R=198$~mm. The radial distributions of signal and background \Stwo signals are shown in the left panel of Figure~\ref{fig:selection_fiducial}, together with the cut position. With this cut we keep 76\% of the simulated signal events and 9\% of the background events.

For the longitudinal coordinate we set both a lower and an upper limit. The optimal values are $Z_{\mathrm{min}}=44.0$~mm and $Z_{\mathrm{max}}=512.5$~mm, respectively. The TPC active volume boundaries along drift are $Z=0$ (anode) and $Z=530.3$~mm (cathode). The drift distance distributions of signal and background \Stwo signals are shown in the right panel of Figure~\ref{fig:selection_fiducial}, together with the cut positions.
Combining the lower and upper edges of the cut we end up keeping 93.5\% of the signal events and 32\% of the background events.

\subsection{Event energy selection}
\label{subsec:selection_energy}

As mentioned in Section~\ref{subsec:data_signal}, the energy released in our simulated \Xe{124} \ecectwonu events is 63.63~keV. Considering the good energy resolution of the detector, it is rather convenient to restrict the energy ROI beyond that already done by the \Kr{83\mathrm{m}} veto requirement of Section~\ref{subsec:selection_krveto}.

\begin{figure}[t]
  \begin{center}
    \includegraphics[width=0.5\textwidth]{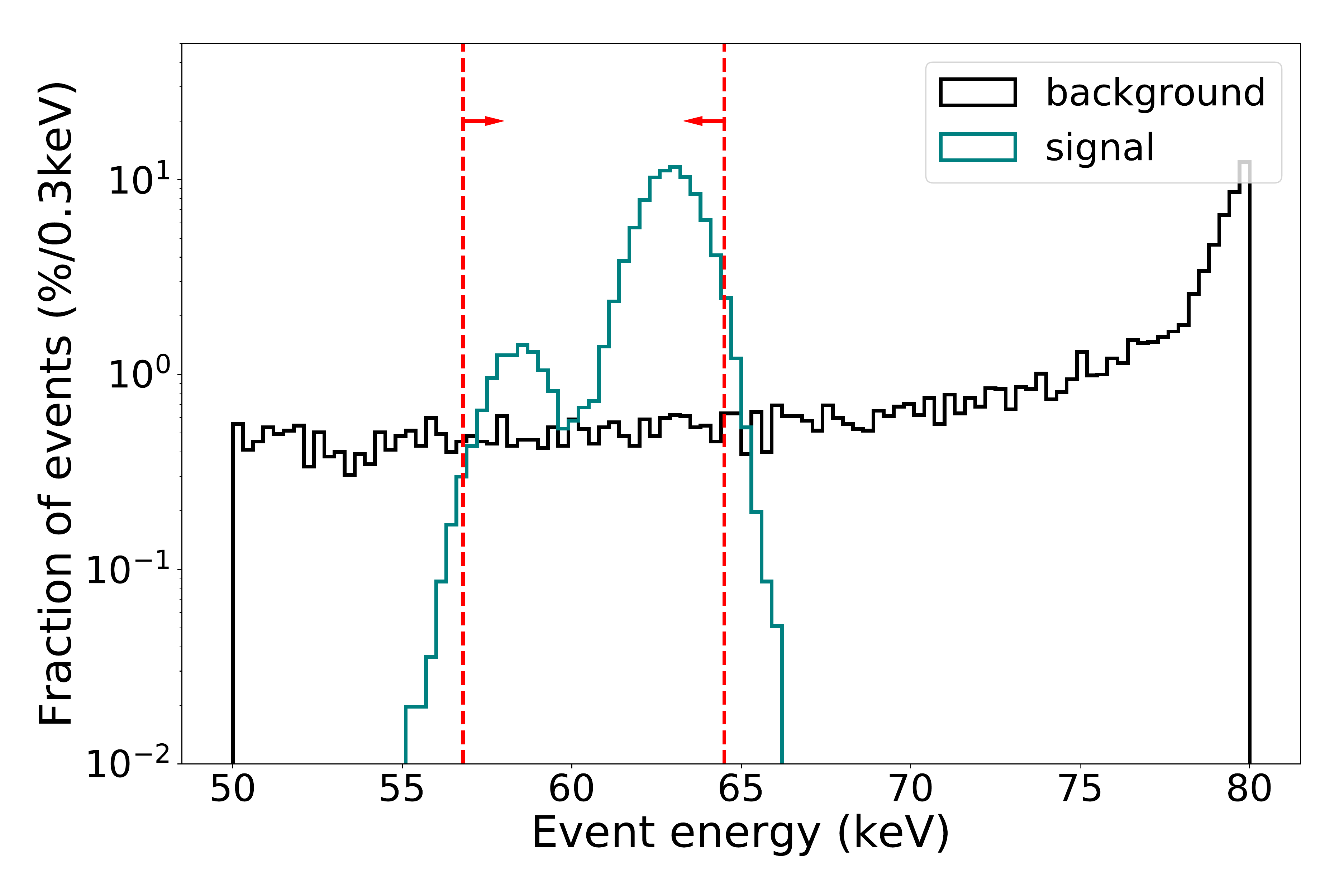}
    \caption{Event energy, $E_{\mathrm{evt}}$, distribution for the signal and background datasets after fiducial selection, with optimal $E_{\mathrm{evt}}$ cut positions (red lines).}
    \label{fig:selection_energy}
  \end{center}
\end{figure}

The maximum and minimum allowed event energies are separately optimized to define the energy ROI $E_{\mathrm{evt},\mathrm{min}} < E_{\mathrm{evt}} < E_{\mathrm{evt},\mathrm{max}}$. The optimal value for the upper limit is found to be $E_{\mathrm{evt},\mathrm{max}}=64.6$~keV, and for the lower limit $E_{\mathrm{evt},\mathrm{min}}=56.4$~keV. The combined requirement of both limits yields a 96\% efficiency for signal events and a 13\% efficiency for background events. Event energy distributions for signal and background fiducial events are shown in Figure~\ref{fig:selection_energy}. As can be  seen in the figure, the cut optimization selects a bimodal distribution for signal events: a primary peak corresponding to the full energy deposit being correctly reconstructed, plus a secondary peak where an energy of 4--5~keV either escapes the active volume or is not reconstructed. As shown in Figure~\ref{fig:mctruth}, X-ray lines of 4--5~keV are indeed expected. Background events are found to have a flat energy distribution in the ROI.

\subsection{Multi-site selection}
\label{subsec:selection_multisite}

We account for two signal sub-samples in our analysis: both a two-\Stwo sub-sample and a three-\Stwo sub-sample. For the two-\Stwo sub-sample not all events are selected, but only those where both energy depositions in the event satisfy a certain energy requirement $E_{S_2,\mathrm{min}}<E_{S_2}<E_{S_2,\mathrm{max}}$. As observed in Figure~\ref{fig:mctruth} (right), for signal we expect most isolated energy deposits to be due to tellurium $K_{\alpha}$ (L$\to$K) X-rays, with a \Stwo energy of 27.5~keV. As shown in the same figure, higher-energy X-rays of 31--32~keV from initial electrons in outer shells are also possible.

\begin{figure}[t]
  \begin{center}
    \includegraphics[width=0.5\textwidth]{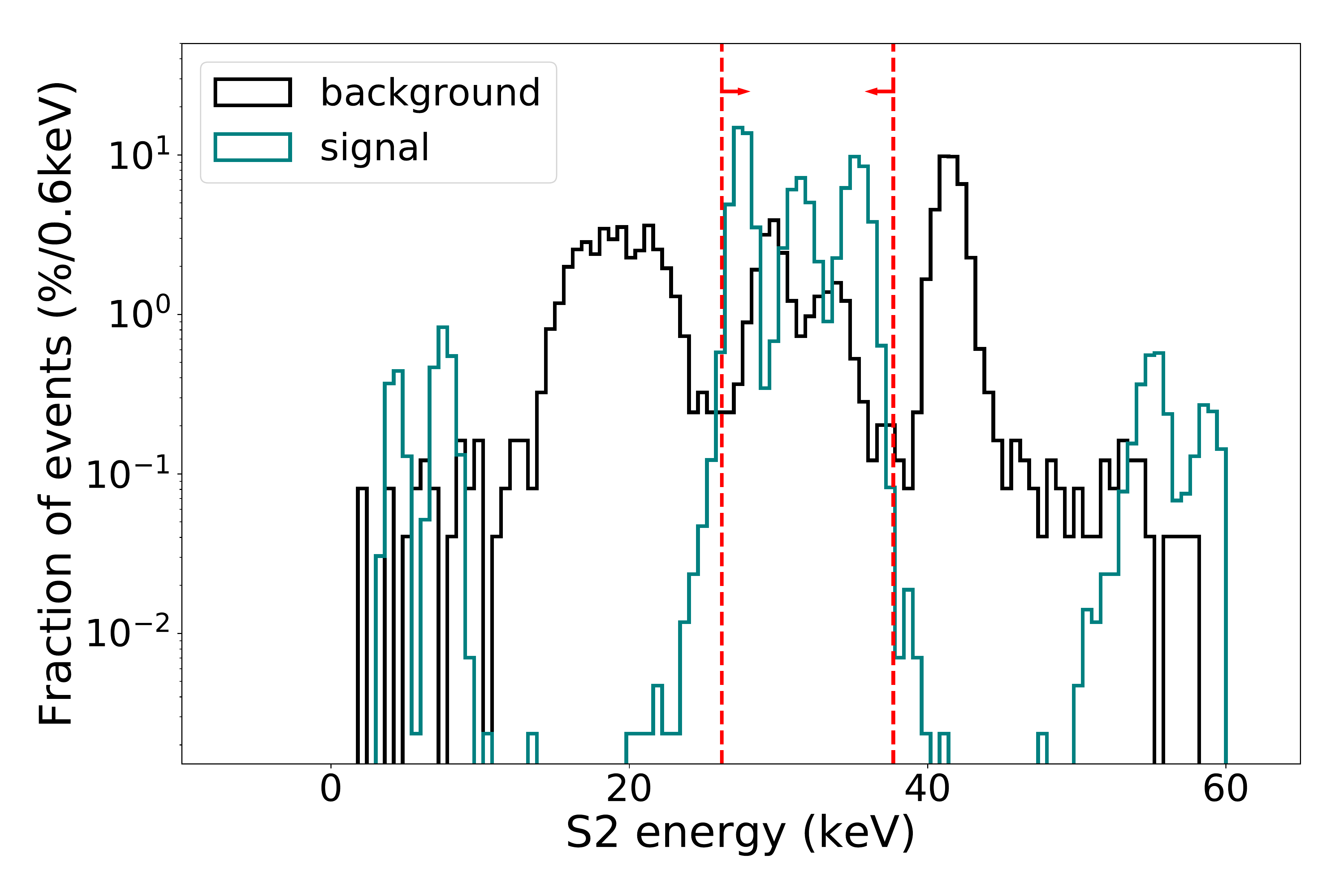}
    \caption{\Stwo energy $E_{S_2}$ signal and background distribution in two-\Stwo sub-sample, with optimal $E_{S_2}$ cut positions (red lines).}
    \label{fig:selection_es2}
  \end{center}
\end{figure}

Figure~\ref{fig:selection_es2} shows the \Stwo energy spectrum of the two-\Stwo signal and background sub-samples. After evaluating the figure of merit we obtain $E_{S_2,\mathrm{min}}=26.0$~keV and $E_{S_2,\mathrm{max}}=37.3$~keV as the optimal limits. The combined effect of both limits keeps 94\% of the signal events and 21.7\% of the background events with 2 \Stwo signals.

As far as the three-\Stwo sub-sample is concerned, no $E_{S_2}$ energy selection is performed and all events in the sub-sample are kept. As shown in Figure~\ref{fig:s2multiplicity}, the probability to have three-\Stwo events is an order of magnitude higher in signal than in background even without a $E_{S_2}$ requirement. After all cuts, the events with three-\Stwo signals represent a 12.7\% of the selected signal events and 0.5\% of the selected background events.


\subsection{Event selection summary}
\label{subsec:selection_summary}

The cumulative efficiencies after each data processing step are shown in Table~\ref{tab:selection} for both background data and signal MC events. We estimate a final efficiency of $(22.90\pm 0.13)$\% for simulated double K~shell captures, to be compared with an acceptance of $(1.15\pm 0.07)\times 10^{-6}$ measured in \Xe{124}-depleted background data. The latter number corresponds to a total background rate of 24.7~$\mu$Hz, or 780~counts/yr.

\begin{figure}[t]
  \begin{center}
    \includegraphics[width=0.5\textwidth]{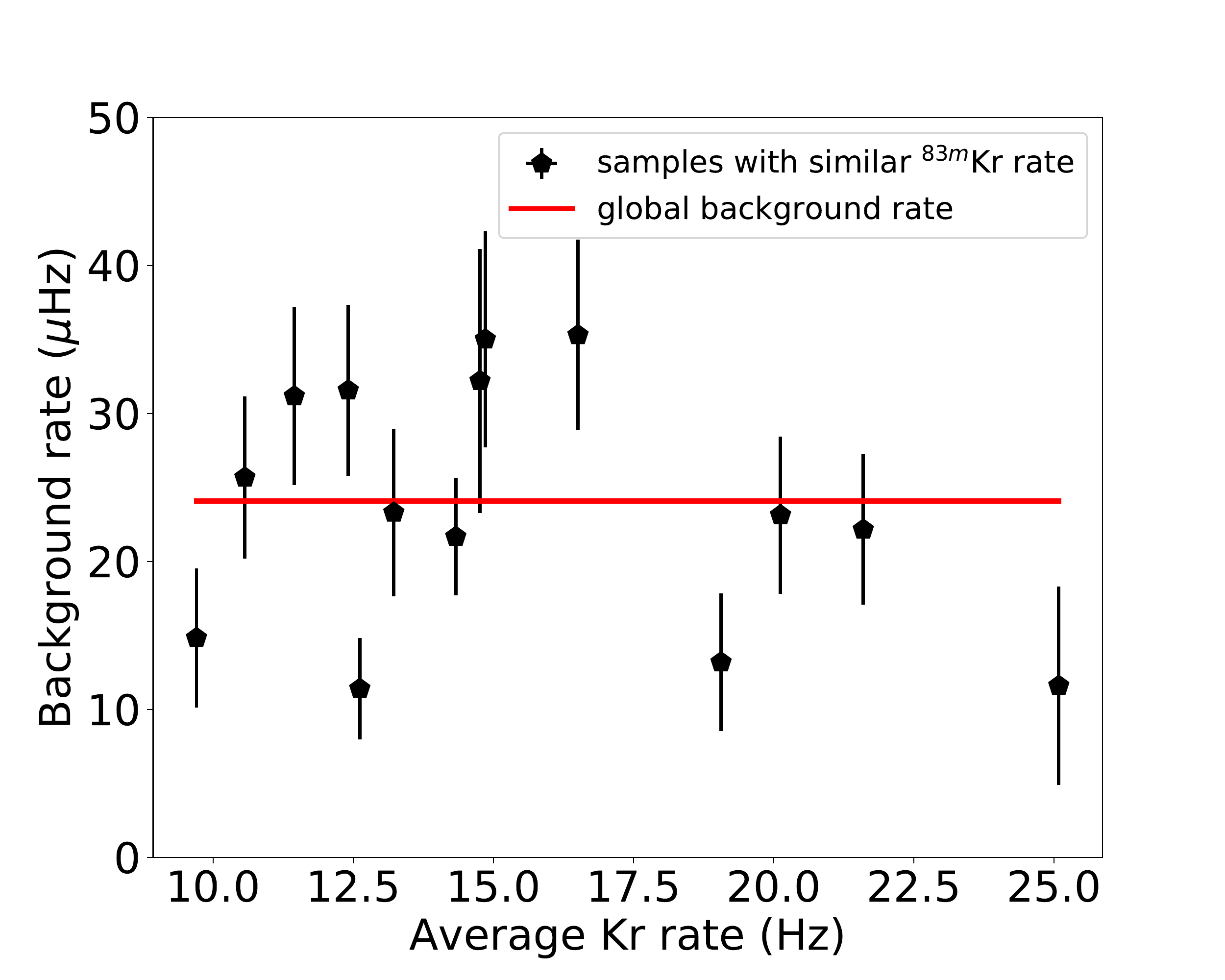}
    \caption{Background rate after all selections as a function of the \Kr{83\mathrm{m}} rate for different samples of the full background dataset. The values oscillate around the overall background rate (solid line).}
    \label{fig:bkgdependence}
  \end{center}
\end{figure}

We also analyze the dependence of the background rate after all selections on the initial rate of events. Since the background sample is dominated by the \Kr{83\mathrm{m}} calibration source, we divide the full dataset into smaller samples with approximately the same \Kr{83\mathrm{m}} rate. The optimized selections from the full dataset are then applied to these samples. Figure~\ref{fig:bkgdependence} shows the background rate after all selections as a function of the average \Kr{83\mathrm{m}} rate in each sample. Clearly, the background rate does not depend on the \Kr{83\mathrm{m}} rate, proving that the selection is robust and can be used in a variety of detector conditions.

While our data-driven approach to the background estimate implies that a full understanding of the background composition is {\it a priori} not known, much can be inferred from the background data distributions. On the one hand, Figure~\ref{fig:bkgdependence} excludes \Kr{83\mathrm{m}}-related events as a dominant background source. On the other hand, Figures~\ref{fig:selection_energy} and \ref{fig:selection_es2} point to multi-site interactions with a flat event energy spectrum in the region of interest, and where one of the energy deposits is a xenon K-shell X-ray at 29.7~keV or 33.8~keV. Background events are therefore likely dominated by low-energy gamma rays produced by radioactive impurities in the detector materials, undergoing photoelectric absorption or by low-angle Compton scattering of higher energy gammas. We cannot determine which case is more likely to happen neither be certain about which gamma-ray emitting isotope is primarily responsible for the overall background rate. Based on findings from other experiments \cite{Abe:2018gyq,XENON:2019dti}  as well as from NEXT measurements at higher energies \cite{Novella:2019cne}, \Co{60}, \K{40}, and isotopes in the \U{238} and \Th{232} decay chains are the likely main contributors.

\section{Sensitivity and precision projections}
\label{sec:sensitivity}

The data-driven background evaluation and the corresponding \ecectwonu signal efficiency study of Section~\ref{sec:selection} have been done using \new data. However, our purpose is to study the feasibility of detecting K~shell \ecectwonu events in the forthcoming \nexthundred detector (see Section~\ref{sec:next}), if a significant \Xe{124} mass were added to its \Xe{136}-enriched xenon gas. The \nexthundred detector is designed to hold approximately 100~kg of xenon with an enrichment fraction of $90\%$ in the \Xe{136} isotope. The natural abundance of \Xe{124} is 0.095\%, or about 100~g of \Xe{124} in 100~kg of natural xenon. In the \Xe{136}-enriched xenon to be used in \nexthundred, the amount of \Xe{124} mass would be even less, as shown in Figure~\ref{fig:xenon}. Such \Xe{124} masses are too small for a competitive \Xe{124} \ecectwonu search. The objective of this study is therefore studying whether the addition of about 1~kg of highly \Xe{124}-enriched xenon to the 100~kg of \Xe{136}-enriched gas could yield a promising \ecectwonu sensitivity in \nexthundred. In the following, we assume that \nexthundred will contain 1~kg of \Xe{124} in its active volume. The mixing of \Xe{136} and \Xe{124} would not affect the performance of the detector in any way, nor would it impact the program of \Xe{136} \bb searches in \nexthundred. The \Xe{136} mass in the active volume would remain essentially  the same  after mixing this relatively small amount of \Xe{124}. The NEXT gas system is already prepared to perform such mixing operation, if a \Xe{124} gas bottle were acquired.

The sensitivity to the \Xe{124} \ecectwonu half-life $T_{1/2}$, where both electrons are captured from the K~shell, is computed as \cite{GomezCadenas:2011it}:

\begin{equation}
T_{1/2} =\ln 2\cdot \frac{ N_{A}\cdot \varepsilon_{s}\cdot M\cdot t}{W\cdot N_{s}}
\label{eq:sensitivityformula}
\end{equation}

\noindent where $N_{A}$ the Avogadro's number, $\varepsilon_{s}=(22.90\pm 0.13)$\% is the \ecectwonu signal efficiency estimated in Section~\ref{sec:selection}, $M$ is the \Xe{124} mass in the \nexthundred active volume, $t$ is the exposure time, $W=123.9$~g/mol is the molar mass of the \Xe{124} isotope, and $N_{s}$ is the maximum number of \ecectwonu signal events that would be compatible with a background-only measurement. Note that the average upper limit $N_{s}$ depends on the number of background events for an exposure time $t$, which has been estimated in Section~\ref{sec:selection}, and on the confidence level value assumed. In the following, we use the customary 90\% confidence level value. There are various prescriptions to evaluate $N_{s}$ given a certain background prediction. In this case, we use the standard Feldman-Cousins prescription \cite{Feldman:1997qc}. The \ecectwonu sensitivity of Eq.~(\ref{eq:sensitivityformula}) refers specifically to the double K~shell capture process in \Xe{124}, as the event selection would be different for different capture configurations.

The XENON1T measurement yields a relatively high, 4.4$\sigma$, evidence for \Xe{124} \ecectwonu. Therefore, in the following we also compute NEXT statistical precision in measuring the half-life assuming that XENON1T extracted central value is the true half-life of the process. Further assuming that the background prediction is perfectly known, and hence the only error in the signal extraction comes from the Poisson fluctuation in the total number of observed events, the relative statistical precision on the half-life for a given exposure can simply be written as:

\begin{equation}
\delta T_{1/2}/T_{1/2} = \delta S/S = \sqrt{S+B}/S
\label{eq:precision}
\end{equation}

\noindent where $B$ and $S$ are the number of background and signal events passing all cuts, respectively.


\subsection{Nominal sensitivity and precision}
\label{subsec:sensitivity_nominal}

\begin{figure}[t]
  \begin{center}
    \includegraphics[width=0.7\textwidth]{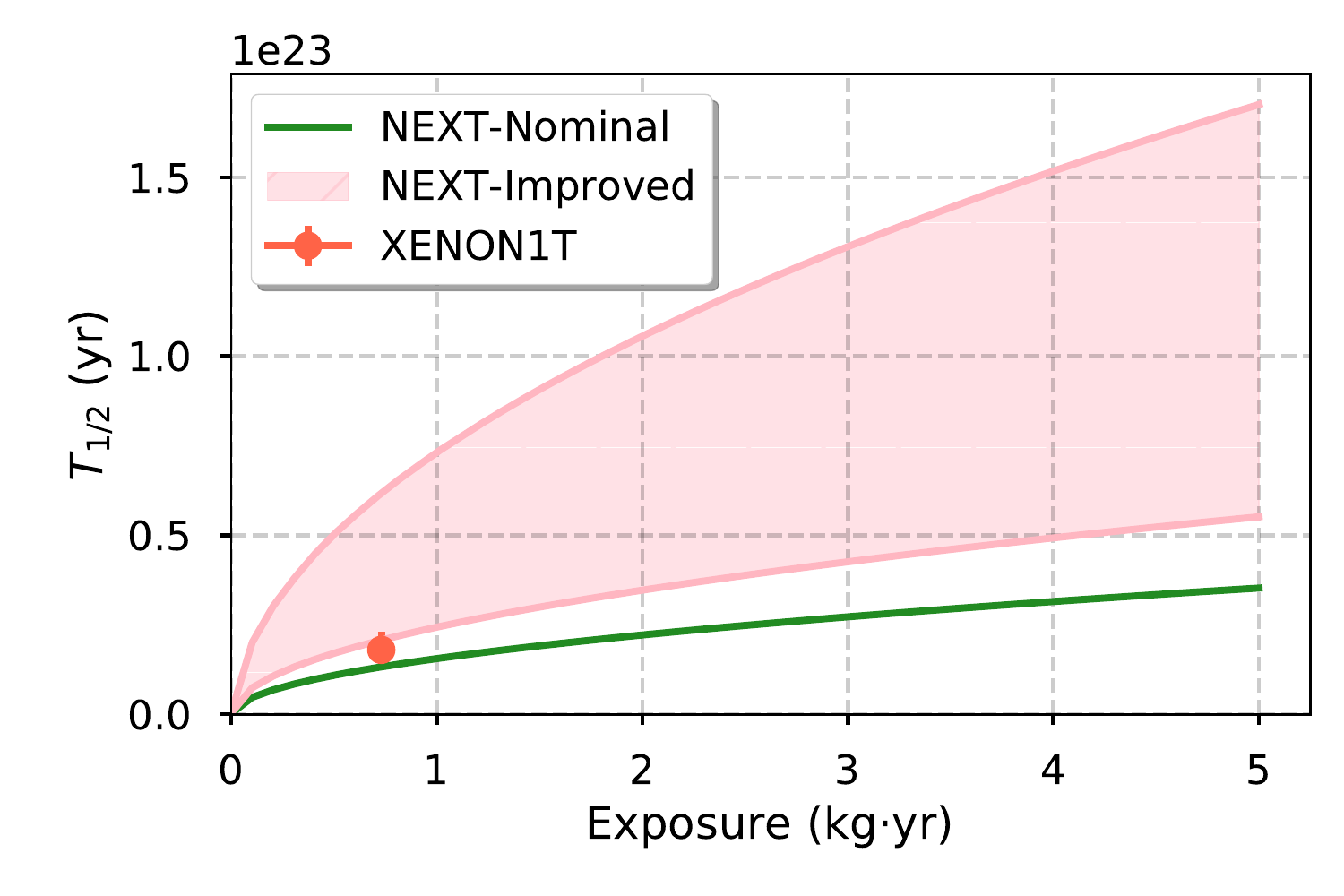}
    \caption{NEXT nominal (green curve) and improved (red band) half-life sensitivity to \Xe{124} \ecectwonu versus exposure, compared with the XENON1T measurement in reference  \cite{XENON:2019dti} (red marker).}
    \label{fig:sensitivity}
  \end{center}
\end{figure}

The NEXT half-life sensitivity to \Xe{124} \ecectwonu is shown as a function of accumulated exposure (in kg$\cdot$yr) as the green line in Figure~\ref{fig:sensitivity}. A sensitivity of $1.6\times 10^{22}$~yr at 90\% CL, comparable to the central value of the recent XENON1T measurement \cite{XENON:2019dti}, is expected to be reached after an exposure of 1~kg$\cdot$yr in \nexthundred. This is one year livetime of \nexthundred operated with 1~kg of \Xe{124} mass in its active volume. In five years of operation in the same conditions, a sensitivity of $3.5\times 10^{22}$~yr at 90\% CL could be reached. We consider this nominal sensitivity to be conservative, as it assumes the background rate measured in \new and the \ecectwonu signal efficiency achieved in this work, as presented in Section~\ref{sec:selection}.

\begin{figure}[t]
  \begin{center}
    \includegraphics[width=0.7\textwidth]{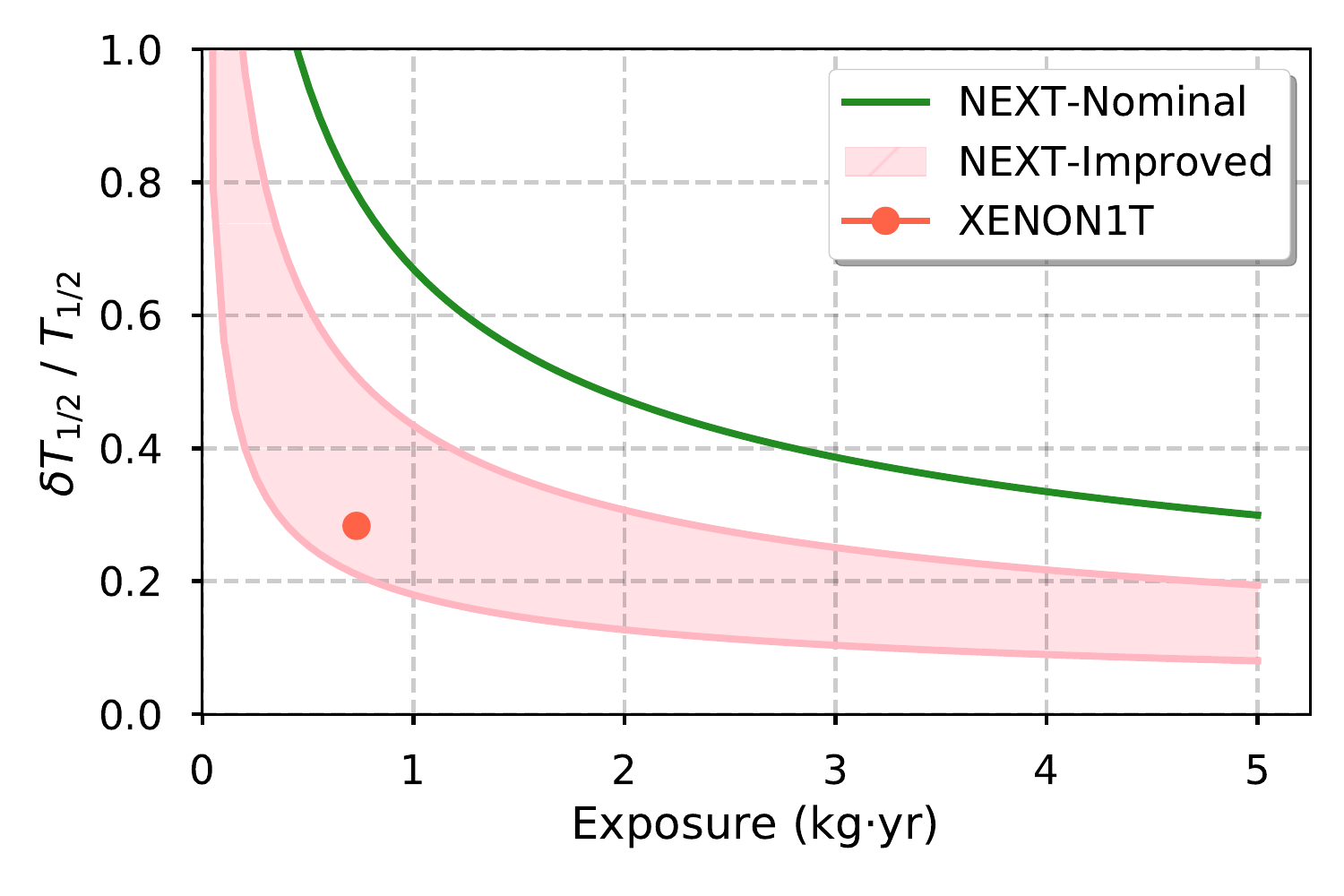}
    \caption{NEXT nominal (green curve) and improved (red band) relative precision in the measurement of the \Xe{124} \ecectwonu half-life as a function of exposure, assuming a true value of $\halflife = 1.8\times 10^{22}$~yr from \cite{XENON:2019dti}. The red marker shows the precision accomplished in \cite{XENON:2019dti}.}
    \label{fig:precision}
  \end{center}
\end{figure}

NEXT relative statistical precision in the \Xe{124} \ecectwonu half-life measurement as a function of exposure, assuming a true value of $\halflife = 1.8\times 10^{22}$~yr, is shown as the green  curve in Figure~\ref{fig:precision}. For such nominal background rate and signal efficiency assumptions, a 30\% precision is expected after a 5~kg$\cdot$yr exposure.

In the following, we justify how more favorable assumptions for background rate and signal efficiency are plausible for \nexthundred, and we quantify their impact on the NEXT sensitivity and precision to double electron capture.


\subsection{Potential improvements}
\label{subsec:sensitivity_improved}

The first potential improvement going from \new to \nexthundred is the size of the detector. As the radius and length of the TPC are increased, the fiducial regions described in Section~\ref{subsec:selection_fiducial} represent a larger fraction of the active volume. Thus, a larger fraction of the \Xe{124} mass is kept, which translates into an increase in signal rate.

The second improvement comes from changes in detector design that reduce the background rate. Compared to \new, \nexthundred is being built from more radiopure materials and with a thicker inner shielding. In \nexthundred, the inner shielding is made of 12~cm of ultra-pure copper, to be compared with the 6~cm copper shielding of \new. The thicker shielding, together with the larger active volume, is expected to be particularly important to mitigate the low-energy ($\lesssim$100~keV) backgrounds relevant to \ecectwonu searches. 

\begin{figure}[t]
  \begin{center}
    \includegraphics[width=0.49\textwidth]{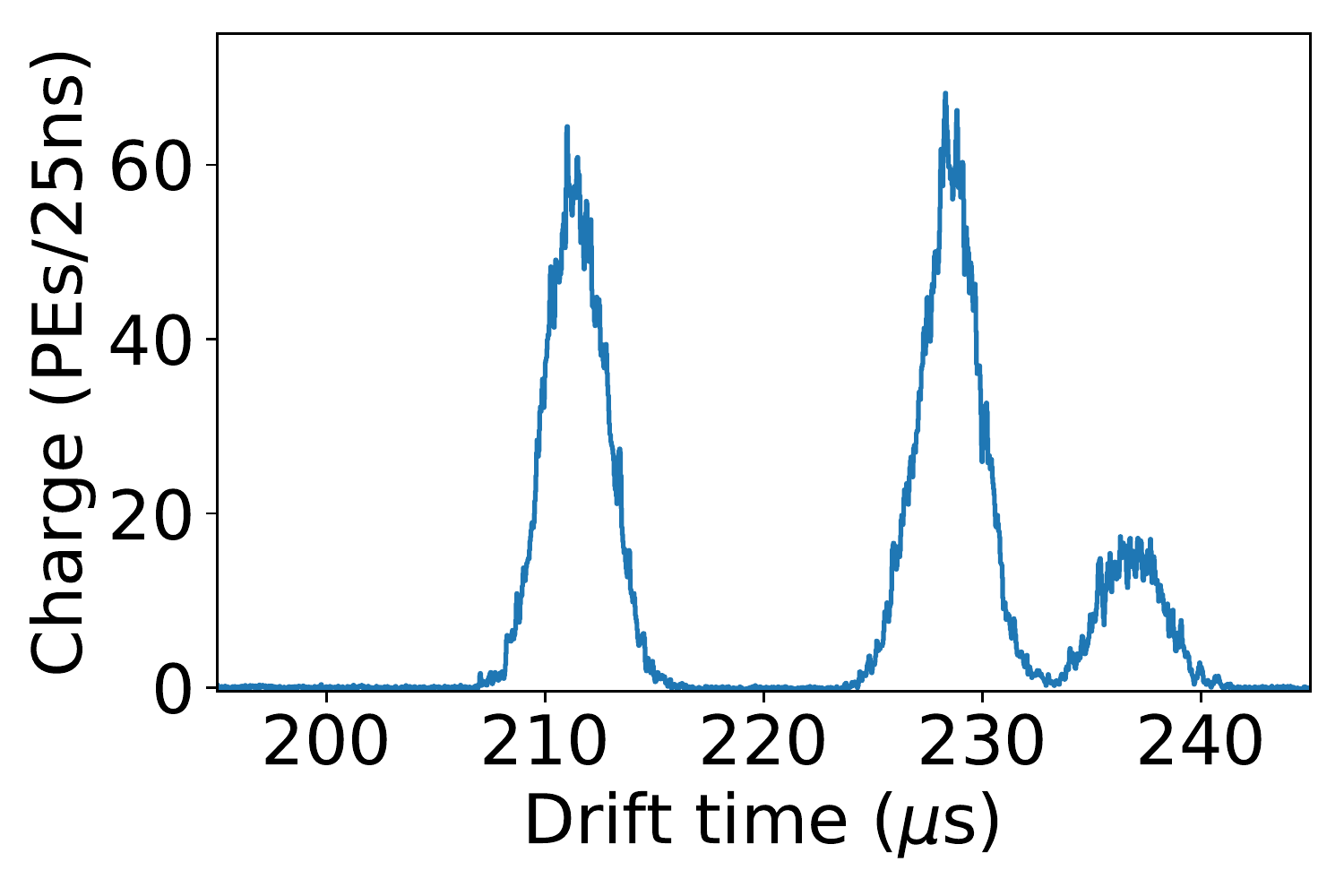} \hfill
    \includegraphics[width=0.49\textwidth]{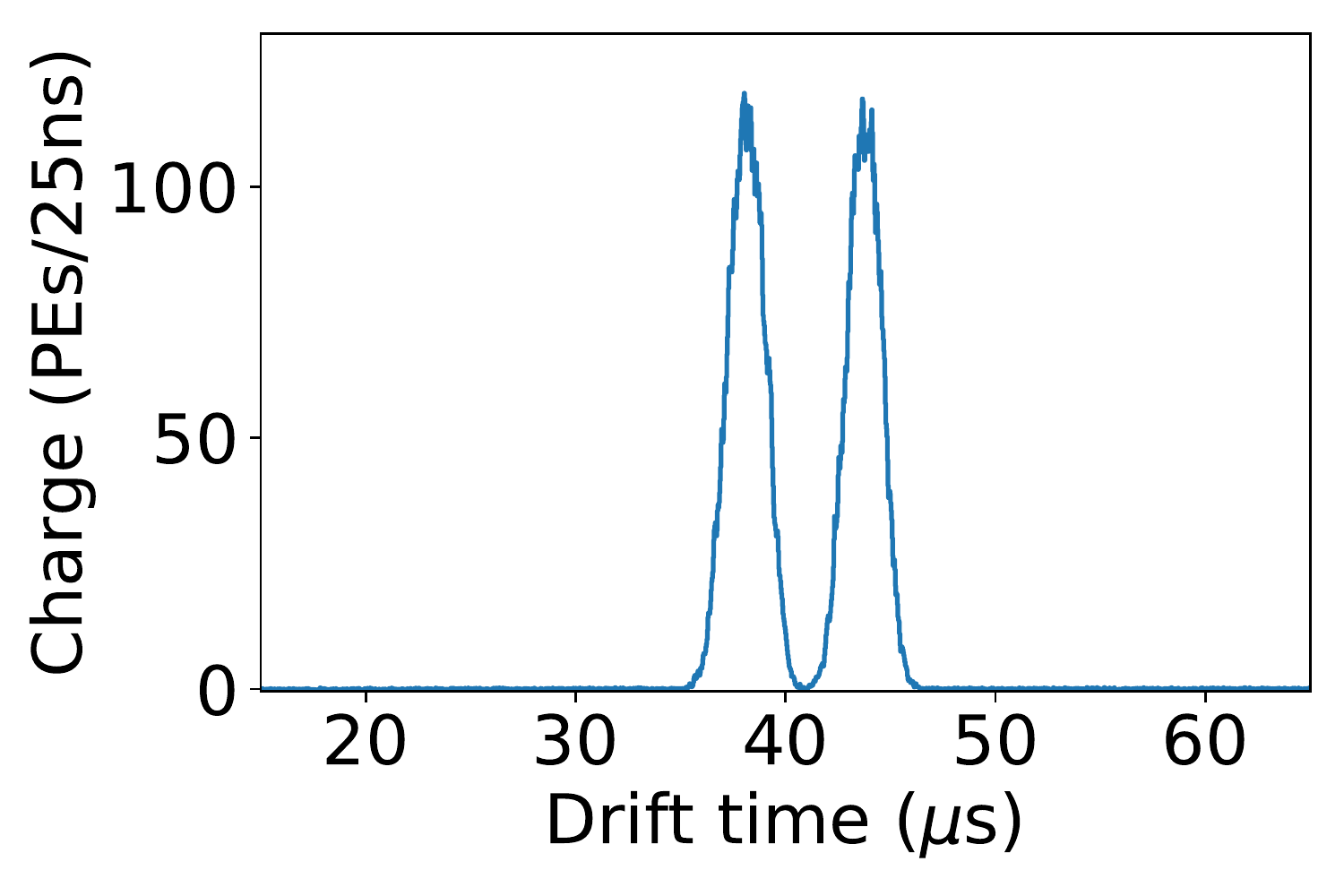}
    \caption{Examples of PMT waveforms of simulated \ecectwonu signal events that are incorrectly identified by NEXT standard reconstruction algorithms.  Left panel: event with three separate energy deposits, reconstructed as having two \Stwo signals at times 211 and 228~$\mu$s, respectively. Right panel: event with two separate energy deposits, reconstructed as having one \Stwo signal at time 38~$\mu$s.}
    \label{fig:waveforms}
  \end{center}
\end{figure}

Third, improvements in low-energy event reconstruction should significantly increase the \ecectwonu signal efficiency in \nexthundred, and possibly also reduce the background rate. NEXT reconstruction algorithms used in Section~\ref{sec:data} and Section~\ref{sec:selection} were developed for the higher-energy \bb searches. Dedicated low-energy reconstruction algorithms are likely to perform better. We show in Figure~\ref{fig:waveforms} examples of PMT waveforms for simulated \ecectwonu events that are incorrectly reconstructed by the current algorithms. As it is apparent from the waveforms, the reconstruction tends to merge into the same \Stwo signal energy deposits that are visually separable from each other. This is due to the proximity of energy deposits along the drift direction and to the longitudinal diffusion, partially merging the nearby deposits together. As noted earlier (Section~\ref{sec:next}), spatial separations of order 10--20~mm among energy  deposits in the same event are common, in the \ecectwonu case. In particular, a large inefficiency results from signal events that are mis-reconstructed as single \Stwo signal events, which are discarded as background-like in our analysis. In addition to more sophisticated hit finding strategies, the performance of recently developed deconvolution algorithms indicates that the effect of electron diffusion can be reduced to a negligible level, which would also improve our ability to resolve close-by energy depositions. It is therefore plausible to obtain a large overall signal efficiency thanks to improvements in the reconstruction procedures combined with the larger active volume of \nexthundred.

Fourth, event selection might also be improved in \nexthundred. One possibility would be to adopt a multi-variate selection as opposed to sequential cuts as done in Section~\ref{sec:selection}. In this case the selection would be optimized in a multi-dimensional space of observables, for example in ($R$, $Z$, $E_{\mathrm{evt}}$, $E_{S_2}$) space, accounting for correlations among them. On the other hand, another possibility to improve the signal/background event classification is by using deep learning techniques. This technique is already being used in NEXT for \bbnonu searches, yielding promising results \cite{Renner:2016tr}.

We therefore believe that an improved half-life sensitivity in \nexthundred is plausible, compared to the nominal sensitivity obtained from \new detector performance and backgrounds. In order to quantify this improvement, we provide an estimate of the signal efficiency assuming the optimal selection cuts described in Section \ref{sec:selection} applied on a simulated dataset of \ecectwonu events in the \nexthundred (as opposed to \new) detector geometry with negligible diffusion. In this baseline scenario, we obtain an improved signal efficiency of 35.82\%, to be compared with the nominal one of 22.90\% in Table~\ref{tab:selection}. The other major contributor to the sensitivity, the background rate, cannot be determined precisely. Nonetheless, a low-energy background rate reduction compared to \new is likely, as motivated above. Therefore for our optimistic scenario, we assume up to a factor-10 improvement on the background rate from the \new detector on top of the signal efficiency improvement.

This improved sensitivity is shown in Figure~\ref{fig:sensitivity} as a filled red band. The lower limit of the band is the baseline scenario with increased \ecectwonu efficiency, $\varepsilon_s$, and minimal diffusion. The upper limit of the band assumes a factor of 10 reduction in background rate on top of that. Sensitivities in the $0.6$--$1.7\times 10^{23}$~yr range at 90\% CL and after 5 years of operation appear possible.

The improved assumptions on \ecectwonu efficiency and background rate have also been used to quantify NEXT half-life measurement precision in a more optimistic scenario compared to Section~\ref{subsec:sensitivity_nominal}. As shown by the red band in Figure~\ref{fig:precision}, a significantly better precision in the 8-19\% range would be achievable after a 5~kg$\cdot$yr exposure. We also note that upcoming multi-tonne liquid xenon time projection chambers for direct dark matter detection, such as XENONnT \cite{Aprile:2020vtw}, LZ \cite{Akerib:2019fml} or PandaX-4T \cite{Zhang:2018xdp},  should also be able to significantly improve the sensitivity and precision reach for \Xe{124} \ecectwonu, compared to XENON1T. For a detailed discussion of the detection prospects for the second-order weak decays of \Xe{124} in multi-tonne xenon time projection chambers, we refer to \cite{Wittweg:2020fak}.

While this study focuses on \Xe{124} \ecectwonu, an accurate  measurement of the two-neutrino mode would set the stage for a sensitive search of the lepton number-violating \ececnonu mode of \Xe{124}, which would be the ultimate goal for double electron capture searches in NEXT.

\section{Conclusions}
\label{sec:conclusions}

The potential discovery of the Majorana nature of massive neutrinos via the observation of the neutrinoless decay modes of double beta decay processes is one of the most important questions in neutrino physics today. Neutrinoless double electron capture on proton-rich nuclei is a promising and alternative process compared to the far more exploited neutrinoless double $\beta^-$ decay of neutron-rich nuclei. Yet, this process is significantly less understood from both the theoretical and the experimental points of view. The measurement of the Standard Model-allowed two-neutrino double electron capture is a first important step toward sensitive neutrinoless double electron capture searches. Two-neutrino double electron capture is a process in which two orbital electrons are simultaneously captured, typically from the K~shell, by a proton-rich nucleus. The experimental signature is given by the emission of X-rays and Auger electrons from the de-excitation of the daughter atom.

In this paper, we establish that the high-pressure xenon gas TPC technology developed by the NEXT Collaboration for \Xe{136} double $\beta^-$ decay searches is ideally suited to perform \Xe{124} double electron capture searches as well. The reasons are the excellent energy resolution of the technology at the energy region of interest near 64~keV, its 3D imaging capabilities to suppress external backgrounds, and its capability to spatially separate the X-ray conversions or Auger electron deposits for a significant fraction of all double electron capture events. We have studied the feasibility to detect \Xe{124} two-neutrino double electron capture in NEXT using actual \new background data and a detailed simulation of the signal in the same detector. The low-background data sample uses 125.9~days of \new low-energy triggers, originally designed for \Kr{83\mathrm{m}} detector calibrations, with the detector filled with \Xe{124}-depleted xenon gas ($<10^{-5}$ isotopic abundance). Both background data and simulated signal are reconstructed with NEXT standard reconstruction algorithms. An optimal event selection for \Xe{124} two-neutrino double electron capture searches has been developed, maximizing the signal efficiency over the square root of background acceptance as the figure of merit. The event selection relies on the total event energy, on the multiplicity of xenon primary (\Sone) and charge-induced (\Stwo) scintillation signals per event, and on the spatial location and energy of the individual energy depositions reconstructed in the event. As a result, a total background rate of 24.7~$\mu$Hz (780 counts/yr) is measured in \new, for a total signal efficiency of 22.9\%. Extrapolating this background rate and signal efficiency to \nexthundred and assuming it is operated with 1~kg of \Xe{124} in its active volume, an option that is technically feasible if a bottle with sufficient \Xe{124}-enriched gas quantity were procured, a sensitivity of $1.6\times 10^{22}$~yr at 90\% CL could be obtained after one year of operations. This sensitivity is comparable to the recent measurement of \Xe{124} by the XENON1T Collaboration, $(1.8\pm 0.5)\times 10^{22}$~yr \cite{XENON:2019dti}. We use this result as a basis to assess the case for a \Xe{124} two-neutrino double electron capture measurement in the \nexthundred detector, currently under construction. We expect that a dedicated double electron capture reconstruction capable of removing diffusion effects and better discerning separate energy deposits could increase the total signal efficiency to 35.82\%. With this improvement alone, we predict a two-neutrino double electron capture sensitivity of $6\times 10^{22}$~yr after 5 years of operation, for the same background rate (780 counts/yr) and the same \Xe{124} active mass (1~kg). Assuming that XENON1T extracted central value is the true half-life of the process, a relative statistical precision of 19\% in the half-life measurement is expected in \nexthundred after 5 years. Other \nexthundred detector design and analysis improvements compared to NEXT-White may yield further improvements in background reduction, resulting in even better sensitivity and precision.

\acknowledgments
The NEXT Collaboration acknowledges support from the following agencies and institutions: the European Research Council (ERC) under the Advanced Grant 339787-NEXT; the European Union's Framework Programme for Research and Innovation Horizon 2020 (2014-2020) under the Marie Skłodowska-Curie Grant Agreements No. 674896, 690575 and 740055; the Ministerio de Econom\'ia y Competitividad and the Ministerio de Ciencia, Innovaci\'on y Universidades of Spain under grants FIS2014-53371-C04, RTI2018-095979, the Severo Ochoa Program grants SEV-2014-0398 and CEX2018-000867-S, and the Mar\'ia de Maeztu Program MDM-2016-0692; the GVA of Spain under grants PROMETEO/2016/120 and SEJI/2017/011; the Portuguese FCT under project PTDC/FIS-NUC/2525/2014, under project UID/FIS/04559/2013 to fund the activities of LIBPhys, and under grants PD/BD/105921/2014, SFRH/BPD/109180/2015 and SFRH/BPD/76842/2011; the U.S.\ Department of Energy under contracts number DE-AC02-06CH11357 (Argonne National Laboratory), DE-AC02-07CH11359 (Fermi National Accelerator Laboratory), DE-FG02-13ER42020 (Texas A\&M) and DE-SC0019223 / DE-SC0019054 (University of Texas at Arlington); and the University of Texas at Arlington. DGD acknowledges Ramon y Cajal program (Spain) under contract number RYC-2015-18820. We also warmly acknowledge the Laboratori Nazionali del Gran Sasso (LNGS) and the Dark Side collaboration for their help with TPB coating of various parts of the NEXT-White TPC. Finally, we are grateful to the Laboratorio Subterr\'aneo de Canfranc for hosting and supporting the NEXT experiment.

\bibliographystyle{JHEP}
\bibliography{biblio}

\providecommand{\href}[2]{#2}\begingroup\raggedright\begin{thebibliography}{10}

\bibitem{Fukuda:1998mi}
{\bf Super-Kamiokande} Collaboration, Y.~Fukuda et~al., {\it {Evidence for
  oscillation of atmospheric neutrinos}},  {\em Phys. Rev. Lett.} {\bf 81}
  (1998) 1562--1567, [\href{http://xxx.lanl.gov/abs/hep-ex/9807003}{{\tt
  hep-ex/9807003}}].

\bibitem{Ahmad:2001an}
{\bf SNO} Collaboration, Q.~R. Ahmad et~al., {\it {Measurement of the rate of
  $\nu_e+d \to p+p+e^-$ interactions produced by $^8B$ solar neutrinos at the
  Sudbury Neutrino Observatory}},  {\em Phys. Rev. Lett.} {\bf 87} (2001)
  071301, [\href{http://xxx.lanl.gov/abs/nucl-ex/0106015}{{\tt
  nucl-ex/0106015}}].

\bibitem{Ahmad:2002jz}
{\bf SNO} Collaboration, Q.~R. Ahmad et~al., {\it {Direct evidence for neutrino
  flavor transformation from neutral current interactions in the Sudbury
  Neutrino Observatory}},  {\em Phys. Rev. Lett.} {\bf 89} (2002) 011301,
  [\href{http://xxx.lanl.gov/abs/nucl-ex/0204008}{{\tt nucl-ex/0204008}}].

\bibitem{GomezCadenas:2011it}
J.~J. Gomez-Cadenas, J.~Martin-Albo, M.~Mezzetto, F.~Monrabal, and M.~Sorel,
  {\it {The Search for neutrinoless double beta decay}},  {\em Riv. Nuovo Cim.}
  {\bf 35} (2012) 29--98, [\href{http://xxx.lanl.gov/abs/1109.5515}{{\tt
  arXiv:1109.5515}}].

\bibitem{Winter:1955zz}
R.~G. Winter, {\it {Double K Capture and Single K Capture with Positron
  Emission}},  {\em Phys. Rev.} {\bf 100} (1955) 142--144.

\bibitem{Bernabeu:1983yb}
J.~Bernabeu, A.~De~Rujula, and C.~Jarlskog, {\it {Neutrinoless Double Electron
  Capture as a Tool to Measure the $\nu_e$ Mass}},  {\em Nucl. Phys.} {\bf
  B223} (1983) 15--28.

\bibitem{Aprile:2016qsw}
{\bf XENON} Collaboration, E.~Aprile et~al., {\it {Search for two-neutrino
  double electron capture of $^{124}$Xe with XENON100}},  {\em Phys. Rev.} {\bf
  C95} (2017), no.~2 024605, [\href{http://xxx.lanl.gov/abs/1609.03354}{{\tt
  arXiv:1609.03354}}].

\bibitem{Abe:2018gyq}
{\bf XMASS} Collaboration, K.~Abe et~al., {\it {Improved search for
  two-neutrino double electron capture on $^{124}$Xe and $^{126}$Xe using
  particle identification in XMASS-I}},  {\em PTEP} {\bf 2018} (2018), no.~5
  053D03, [\href{http://xxx.lanl.gov/abs/1801.03251}{{\tt arXiv:1801.03251}}].

\bibitem{Gavriljuk:2018pez}
{\relax Yu}.~M. Gavriljuk, A.~M. Gangapshev, V.~V. Kazalov, V.~V. Kuzminov,
  S.~I. Panasenko, S.~S. Ratkevich, and D.~A. Tekueva, {\it {2K-Capture
  in$^{124}$Xe: Results of Data Processing for an Exposure of 37.7 kg day}},
  {\em Phys. Part. Nucl.} {\bf 49} (2018), no.~4 563--568,
  [\href{http://xxx.lanl.gov/abs/1806.03060}{{\tt arXiv:1806.03060}}].

\bibitem{XENON:2019dti}
{\bf XENON} Collaboration, E.~Aprile et~al., {\it {Observation of two-neutrino
  double electron capture in $^{124}$Xe with XENON1T}},  {\em Nature} {\bf 568}
  (2019), no.~7753 532--535, [\href{http://xxx.lanl.gov/abs/1904.11002}{{\tt
  arXiv:1904.11002}}].

\bibitem{Meshik:2001ra}
A.~P. Meshik, C.~M. Hohenberg, O.~V. Pravdivtseva, and {\relax Ya}.~S. Kapusta,
  {\it {Weak decay of Ba-130 and Ba-132: Geochemical measurements}},  {\em
  Phys. Rev.} {\bf C64} (2001) 035205.

\bibitem{PUJOL20096834}
M.~Pujol, B.~Marty, P.~Burnard, and P.~Philippot, {\it Xenon in archean barite:
  Weak decay of 130ba, mass-dependent isotopic fractionation and implication
  for barite formation},  {\em Geochimica et Cosmochimica Acta} {\bf 73}
  (2009), no.~22 6834 -- 6846.

\bibitem{Ratkevich:2017kaz}
S.~S. Ratkevich, A.~M. Gangapshev, {\relax Yu}.~M. Gavrilyuk, F.~F. Karpeshin,
  V.~V. Kazalov, V.~V. Kuzminov, S.~I. Panasenko, M.~B. Trzhaskovskaya, and
  S.~P. Yakimenko, {\it {Comparative study of the double $K$-shell-vacancy
  production in single- and double-electron capture decay}},  {\em Phys. Rev.}
  {\bf C96} (2017), no.~6 065502,
  [\href{http://xxx.lanl.gov/abs/1707.07171}{{\tt arXiv:1707.07171}}].

\bibitem{10030100097}
J.~LAETER, {\it Atomic weights of the elements : review 2000},  {\em Pure Appl.
  Chem.} {\bf 75} (2003) 683--800.

\bibitem{Suhonen:2013rca}
J.~Suhonen, {\it {Double beta decays of $^{124}$Xe investigated in the QRPA
  framework}},  {\em J. Phys.} {\bf G40} (2013) 075102.

\bibitem{Pirinen:2015sma}
P.~Pirinen and J.~Suhonen, {\it {Systematic approach to $\beta$ and
  2$\nu\beta\beta$ decays of mass $A=$100–136 nuclei}},  {\em Phys. Rev.}
  {\bf C91} (2015), no.~5 054309.

\bibitem{Perez:2018cly}
E.~A. Coello~Pérez, J.~Menéndez, and A.~Schwenk, {\it {Two-neutrino double
  electron capture on $^{124}$Xe based on an effective theory and the nuclear
  shell model}},  {\em Phys. Lett.} {\bf B797} (2019) 134885,
  [\href{http://xxx.lanl.gov/abs/1809.04443}{{\tt arXiv:1809.04443}}].

\bibitem{Doi:1992dm}
M.~Doi and T.~Kotani, {\it {Neutrinoless modes of double beta decay}},  {\em
  Prog. Theor. Phys.} {\bf 89} (1993) 139--160.

\bibitem{Alvarez:2013gxa}
{\bf NEXT} Collaboration, V.~Álvarez et~al., {\it {Operation and first results
  of the NEXT-DEMO prototype using a silicon photomultiplier tracking array}},
  {\em JINST} {\bf 8} (2013) P09011,
  [\href{http://xxx.lanl.gov/abs/1306.0471}{{\tt arXiv:1306.0471}}].

\bibitem{Alvarez:2012yxw}
{\bf NEXT} Collaboration, V.~Alvarez et~al., {\it {Near-Intrinsic Energy
  Resolution for 30 to 662 keV Gamma Rays in a High Pressure Xenon
  Electroluminescent TPC}},  {\em Nucl. Instrum. Meth.} {\bf A708} (2013)
  101--114, [\href{http://xxx.lanl.gov/abs/1211.4474}{{\tt arXiv:1211.4474}}].

\bibitem{Monrabal:2018xlr}
{\bf NEXT} Collaboration, F.~Monrabal et~al., {\it {The Next White (NEW)
  Detector}},  {\em JINST} {\bf 13} (2018), no.~12 P12010,
  [\href{http://xxx.lanl.gov/abs/1804.02409}{{\tt arXiv:1804.02409}}].

\bibitem{Martin-Albo:2015rhw}
{\bf NEXT} Collaboration, J.~Martín-Albo et~al., {\it {Sensitivity of NEXT-100
  to Neutrinoless Double Beta Decay}},  {\em JHEP} {\bf 05} (2016) 159,
  [\href{http://xxx.lanl.gov/abs/1511.09246}{{\tt arXiv:1511.09246}}].

\bibitem{Adams:2020cye}
C.~Adams et~al., {\it {Sensitivity of a tonne-scale NEXT detector for
  neutrinoless double beta decay searches}},
  \href{http://xxx.lanl.gov/abs/2005.06467}{{\tt arXiv:2005.06467}}.

\bibitem{Giuliani:2019uno}
{\bf APPEC Committee} Collaboration, A.~Giuliani, J.~J. Gomez~Cadenas,
  S.~Pascoli, E.~Previtali, R.~Saakyan, K.~Schäffner, and S.~Schönert, {\it
  {Double Beta Decay APPEC Committee Report}},
  \href{http://xxx.lanl.gov/abs/1910.04688}{{\tt arXiv:1910.04688}}.

\bibitem{Nygren:2015xxi}
D.~R. Nygren, {\it {Detecting the barium daughter in $^{136}$Xe
  0-$\nu\beta\beta$ decay using single-molecule fluorescence imaging
  techniques}},  {\em J. Phys. Conf. Ser.} {\bf 650} (2015), no.~1 012002.

\bibitem{Jones:2016qiq}
B.~Jones, A.~McDonald, and D.~Nygren, {\it {Single Molecule Fluorescence
  Imaging as a Technique for Barium Tagging in Neutrinoless Double Beta
  Decay}},  {\em JINST} {\bf 11} (2016), no.~12 P12011,
  [\href{http://xxx.lanl.gov/abs/1609.04019}{{\tt arXiv:1609.04019}}].

\bibitem{McDonald:2017izm}
A.~McDonald et~al., {\it {Demonstration of Single Barium Ion Sensitivity for
  Neutrinoless Double Beta Decay using Single Molecule Fluorescence Imaging}},
  {\em Phys. Rev. Lett.} {\bf 120} (2018), no.~13 132504,
  [\href{http://xxx.lanl.gov/abs/1711.04782}{{\tt arXiv:1711.04782}}].

\bibitem{Thapa:2019zjk}
P.~Thapa, I.~Arnquist, N.~Byrnes, A.~Denisenko, F.~Foss, B.~Jones, A.~Mcdonald,
  D.~Nygren, and K.~Woodruff, {\it {Barium Chemosensors with Dry-Phase
  Fluorescence for Neutrinoless Double Beta Decay}},  {\em Sci. Rep.} {\bf 9}
  (2019), no.~1 15097, [\href{http://xxx.lanl.gov/abs/1904.05901}{{\tt
  arXiv:1904.05901}}].

\bibitem{Rivilla:2019vzd}
I.~Rivilla et~al., {\it {Towards a background-free neutrinoless double beta
  decay experiment based on a fluorescent bicolor sensor}},
  \href{http://xxx.lanl.gov/abs/1909.02782}{{\tt arXiv:1909.02782}}.

\bibitem{Simon:2018vep}
{\bf NEXT} Collaboration, A.~Simón et~al., {\it {Electron drift properties in
  high pressure gaseous xenon}},  {\em JINST} {\bf 13} (2018), no.~07 P07013,
  [\href{http://xxx.lanl.gov/abs/1804.01680}{{\tt arXiv:1804.01680}}].

\bibitem{Martinez-Lema:2018ibw}
{\bf NEXT} Collaboration, G.~Martínez-Lema et~al., {\it {Calibration of the
  NEXT-White detector using $^{83m}\mathrm{Kr}$ decays}},  {\em JINST} {\bf 13}
  (2018), no.~10 P10014, [\href{http://xxx.lanl.gov/abs/1804.01780}{{\tt
  arXiv:1804.01780}}].

\bibitem{Agostinelli:2002hh}
{\bf GEANT4} Collaboration, S.~Agostinelli et~al., {\it {GEANT4: A Simulation
  toolkit}},  {\em Nucl. Instrum. Meth.} {\bf A506} (2003) 250--303.

\bibitem{Perkins:236347}
S.~T. Perkins, M.~H. Chen, D.~E. Cullen, and J.~H. Hubbell, {\em {Tables and
  graphs of atomic subshell and relaxation data derived from the LLNL Evaluated
  Atomic Data Library (EADL), Z=1-100}}.
\newblock Lawrence Livermore Nat. Lab., Livermore, CA, 1991.

\bibitem{RevModPhys.39.125}
J.~A. Bearden and A.~F. Burr, {\it Reevaluation of x-ray atomic energy levels},
   {\em Rev. Mod. Phys.} {\bf 39} (Jan, 1967) 125--142.

\bibitem{PhysRevC.86.044313}
D.~A. Nesterenko, K.~Blaum, M.~Block, C.~Droese, S.~Eliseev, F.~Herfurth,
  E.~Minaya~Ramirez, Y.~N. Novikov, L.~Schweikhard, V.~M. Shabaev, M.~V.
  Smirnov, I.~I. Tupitsyn, K.~Zuber, and N.~A. Zubova, {\it
  Double-$\ensuremath{\beta}$ transformations in isobaric triplets with mass
  numbers $\mathbf{A}=\mathbf{124}$, 130, and 136},  {\em Phys. Rev. C} {\bf
  86} (Oct, 2012) 044313.

\bibitem{Firestone.1997}
R.~Firestone, V.~Shirley, C.~Baglin, S.~Chu, and J.~Zipkin, {\em Table of
  Isotopes, 8th edition}.
\newblock Springer, 1997.

\bibitem{Renner:2019pfe}
{\bf NEXT} Collaboration, J.~Renner et~al., {\it {Energy calibration of the
  NEXT-White detector with 1\% resolution near Q$_{\beta \beta}$
  of$^{136}$Xe}},  {\em JHEP} {\bf 10} (2019) 230,
  [\href{http://xxx.lanl.gov/abs/1905.13110}{{\tt arXiv:1905.13110}}].

\bibitem{Novella:2019cne}
{\bf NEXT} Collaboration, P.~Novella et~al., {\it {Radiogenic Backgrounds in
  the NEXT Double Beta Decay Experiment}},  {\em JHEP} {\bf 10} (2019) 051,
  [\href{http://xxx.lanl.gov/abs/1905.13625}{{\tt arXiv:1905.13625}}].

\bibitem{Feldman:1997qc}
G.~J. Feldman and R.~D. Cousins, {\it {A Unified approach to the classical
  statistical analysis of small signals}},  {\em Phys. Rev.} {\bf D57} (1998)
  3873--3889, [\href{http://xxx.lanl.gov/abs/physics/9711021}{{\tt
  physics/9711021}}].

\bibitem{Renner:2016tr}
{\bf NEXT} Collaboration, J.~Renner et~al., {\it {Background rejection in NEXT
  using deep neural networks}},  {\em JINST} {\bf 12} (2017), no.~01 T01004,
  [\href{http://xxx.lanl.gov/abs/1609.06202}{{\tt arXiv:1609.06202}}].

\bibitem{Aprile:2020vtw}
{\bf XENON} Collaboration, E.~Aprile et~al., {\it {Projected WIMP Sensitivity
  of the XENONnT Dark Matter Experiment}},
  \href{http://xxx.lanl.gov/abs/2007.08796}{{\tt arXiv:2007.08796}}.

\bibitem{Akerib:2019fml}
{\bf LZ} Collaboration, D.~Akerib et~al., {\it {The LUX-ZEPLIN (LZ)
  Experiment}},  {\em Nucl. Instrum. Meth. A} {\bf 953} (2020) 163047,
  [\href{http://xxx.lanl.gov/abs/1910.09124}{{\tt arXiv:1910.09124}}].

\bibitem{Zhang:2018xdp}
{\bf PandaX} Collaboration, H.~Zhang et~al., {\it {Dark matter direct search
  sensitivity of the PandaX-4T experiment}},  {\em Sci. China Phys. Mech.
  Astron.} {\bf 62} (2019), no.~3 31011,
  [\href{http://xxx.lanl.gov/abs/1806.02229}{{\tt arXiv:1806.02229}}].

\bibitem{Wittweg:2020fak}
C.~Wittweg, B.~Lenardo, A.~Fieguth, and C.~Weinheimer, {\it {Detection
  prospects for the second-order weak decays of $^{124}$Xe in multi-tonne xenon
  time projection chambers}},  \href{http://xxx.lanl.gov/abs/2002.04239}{{\tt
  arXiv:2002.04239}}.

\end{thebibliography}\endgroup

\end{document}